\documentclass[]{jfm}
\usepackage{graphicx}
\usepackage{epstopdf, epsfig}
\usepackage{amsmath}
\usepackage{multicol}
\usepackage{empheq}
\usepackage{float}
\usepackage{xcolor}
\usepackage{lineno}

\shorttitle{Guidelines for authors}
\shortauthor{R. Rubini, D. Lasagna and A. Da Ronch}

\title{$l_1$-based sparsification of energy interactions \textcolor{black}{in unsteady lid-driven cavity flow}}

\author{Riccardo Rubini\aff{1}, 
  Davide Lasagna\aff{1} \corresp{\email{davide.lasagna@soton.ac.uk}}
 \and Andrea Da Ronch\aff{1}}

\affiliation{\aff{1}Faculty of Engineering and Physical Sciences, University of Southampton, SO17 1BJ, Southampton, United Kingdom}

\begin{document}

\maketitle

\begin{abstract}

In this paper, sparsity-promoting regression techniques are employed to automatically identify from data relevant triadic interactions between modal structures in large Galerkin-based models of two-dimensional \textcolor{black}{unsteady} flows. The approach produces \textcolor{black}{interpretable,} sparsely-connected models that reproduce the original dynamical behaviour at a much lower computational cost, as fewer triadic interactions need to be evaluated. The key feature of the approach is that dominant interactions are selected systematically from the solution of a convex optimisation problem, with a unique solution, and no \textit{a priori} assumptions on the structure of scale interactions are required. We demonstrate this approach on models of two-dimensional lid-driven cavity flow at Reynolds number $Re = 2 \times 10^4$, where fluid motion is chaotic. To understand the role of the subspace utilised for the Galerkin projection on sparsity characteristics, we consider two families of models obtained from two different modal decomposition techniques. The first uses energy-optimal Proper Orthogonal Decomposition modes, while the second uses modes oscillating at a single frequency obtained from Discrete Fourier Transform of the \textcolor{black}{flow} snapshots. We show that, in both cases, and despite no \textit{a-priori} physical knowledge is incorporated into the approach, relevant interactions across the hierarchy of modes are identified in agreement with the expected picture of scale interactions in two-dimensional turbulence. Yet, substantial structural changes in the interaction pattern and a quantitatively different sparsity are observed. \textcolor{black}{Finally, although not directly enforced in the procedure, the sparsified models have excellent long-term stability properties and correctly reproduce the spatio-temporal evolution of dominant flow structures in the cavity. }
\end{abstract}



\section{Introduction}
In the classical description of developed turbulent flows \citep{lumley1979computational, pope2001turbulent, jimenez2018coherent}, energy is transferred across the hierarchy of coherent structures via nonlinear triadic interactions. Implicit in this picture is the fact that not all interactions have the same importance, but they occur in preferential patterns. In fact, extensive numerical evidence suggests that the nonlinear interaction pattern among coherent structures is sparse. The evolution of structures at a certain length scale depends predominantly upon a subset of all other structures \citep{kraichnan1971inertial,ohkitani1990nonlocality, brasseur1994interscale} and the influence of interactions with the complementary set of structures can be generally neglected with minor global effects. 

Successful attempts to construct a reduced set of equations that exploit this sparsity have been made in the past, often for canonical geometries where triadic interactions are conveniently examined in Fourier space and using a coarse-grained partitioning of the hierarchy of scales. \citet{laval1999nonlocality} considered two-dimensional homogeneous decaying turbulence and developed a reduced set of coupled partial differential equations governing the evolution of the large and small scales. In this model, only dominant terms were retained based on observations from direct numerical simulation. With the goal of identifying fundamental mechanisms underlying wall turbulence, \citet{thomas2015minimal} developed nonlinear reduced models of plane Couette flow directly from the governing equations by first partitioning the flow into a streamwise-averaged mean and a perturbation field, and then neglecting nonlinear interactions among the streamwise varying perturbations, i.e. the perturbation-perturbation nonlinearity \citep{thomas2014self}. The models captured well-established roll-streak dynamical features of \textcolor{black}{wall} turbulence and its statistics in a computationally efficient framework. The models also sustained turbulent dynamics down to minimal configurations where interactions between the streamwise mean flow and only one single streamwise wavenumber are retained.



When reduced-order dynamical representations are derived using Galerkin projection on a low-dimensional subspace identified by a set of modal structures \citep{Fletcher:1984wv,rowley2017model}, triadic interactions are conveniently studied in modal space by examining a third-order coefficient tensor arising from projection of the basis function on the convective term of the Navier-Stokes equations \citep{noack2008finite, noack2011reduced}. Sparsity characteristics have also been observed in this reduced-order setting. \citet{couplet2003intermodal} constructed Galerkin models of the separated turbulent flow past a backward-facing step using Proper Orthogonal Decomposition (POD) modes \citep{lumley1970stochastic,sirovich1987turbulence} and observed that the energy transfer pattern in modal space shares many properties with its counterpart in isotropic homogeneous three-dimensional turbulent flows \citep{yeung}. For instance, the authors observed that interactions are local in modal space and that a direct energy cascade exists. Analogously, \citet{rempfer1994dynamics} examined the power budget of POD modes in a transitional boundary layer and observed that interactions in modal space occur predominantly between triads of modes whose sum of modal indices is equal to zero, similar to energy interactions between Fourier modes in homogeneous turbulence. 
However, classical model order reduction techniques \citep{rowley2017model} have not traditionally exploited this feature. In fact, when modal decompositions such as POD are employed, densely-connected models are usually obtained, as the third-order coefficient tensor is dense \textcolor{black}{(i.e. most coefficients are different from zero)} for inhomogeneous flows without particular symmetries. This hinders the interpretation of the underlying physics of scale\textcolor{black}{s} interactions and increases computational costs, \textcolor{black}{since} all triadic interactions have to be evaluated for \textcolor{black}{advancing} the model forward in time.

The first contribution of this work is that we \textcolor{black}{propose to} apply data-driven techniques \citep{blum1997selection, brunton2016discovering, loiseau2018constrained, brunton2019machine} \textcolor{black}{as a means to} identify relevant triadic interactions in Galerkin \textcolor{black}{models of turbulent flows}. 
\textcolor{black}{Our aim} is to generate reduced order models resolving a wide range of scales while preserving computational efficiency and interpretability by pruning \textcolor{black}{weak} interactions that are not relevant for the dynamics. The cornerstone of the proposed approach is $l_1$-\textcolor{black}{regularised} regression \citep{stat_learning,tibshirani2013lasso}, widely used in the statistical community to extract parsimonious representation of complex datasets containing a subset of predominant features. The non-differentiable, yet convex, nature of the $l_1$ regularisation allows transforming the interaction selection problem into a convex optimisation problem that can be solved efficiently, with \textcolor{black}{a} unique solution. Since no \textit{a priori} knowledge of the dynamics is utilised, the approach is fine-grained and relevant interactions are identified in a mode-by-mode fashion across the hierarchy of modes.
Sparsity-promoting regression techniques have been recently proposed by Brunton and coworkers \citep{brunton2016discovering, Kaiser} in the SINDy framework (Sparse Identification of Nonlinear Dynamics), as a means to discover parsimonious dynamical representations of systems whose underlying (but hidden) evolution equations are sparse in the space of possible functions \citep{brunton2016discovering}. \textcolor{black}{Our work deviates from these efforts in the perspective. When formulated in partial differential form, the Navier-Stokes equations for incompressible flows are indeed structurally sparse, as only few terms -- convection, viscous diffusion and pressure forces to conserve mass -- participate to the overall equilibrium. However, when Galerkin models are derived, such structural sparsity is generally lost. What is preserved is the sparsity in the interaction pattern between scales of motion that emerges \textit{a posteriori} in turbulent realisations. Fundamentally, we aim to exploit this feature and distill a structurally sparse model that reproduces the original behaviour}. In addition, sparsity identification methods have been applied, so far, to relatively small Galerkin models \citep{loiseau2018constrained}, and it is not yet understood if \textcolor{black}{these can be utilised to} identify and extract relevant interactions in larger models in agreement with the established picture of energy interactions in turbulent flows. In this sense, our approach is closer to the recent work of \citet{nair2015network}, \citet{taira2016network} and \citet{nair2017oscillator}. These authors employed network-theoretic sparsification approaches \citep{newman2018networks} to identify key vortex-to-vortex interactions in two-dimensional homogeneous turbulence, obtaining sparse models that capture the essential physics of unsteady fluid flow with a reduced number of interactions between the same large number of states. 

The second contribution of this paper is that we examine how sparsity of energy interactions depends on the subspace used to generate the Galerkin model. 
Finding an appropriate subspace for projection is recognised as a challenging task \citep{Noack:2016jk}, and several modal decompositions have been proposed differing in spirit and approach (see \citet{taira2017modal} for a recent review). However, the role of the subspace on the organisation of energy interactions has not been explored in the past. To address this question, we examine and compare in this paper energy interactions and sparsity features of two families of Galerkin models. The first uses energy-optimal POD modes while the second uses modes oscillating temporally at a single frequency, obtained using a procedure based on Spectral Proper Orthogonal Decomposition \citep{sieber2016spectral} and equivalent to a Discrete Fourier Transform (DFT) of the velocity snapshots. Here, we aim \textcolor{black}{to understand} if the optimal data-representation property of POD also provides the best description in terms of sparsity, even if POD \textcolor{black}{is known to couple flow structures at different spatial or temporal scales} \citep{Noack:2016jk, towne2018spectral}.

This manuscript is organised as follows. For completeness, section \ref{sec:methodology} summarises the methodology utilised to generate reduced order models using Galerkin projection, and then discusses how energy interactions in Galerkin models can be examined. Subsequently, the $l_1$-based sparse regression approach is outlined and conceptual differences between our approach and the SINDy approach proposed in \citet{brunton2016discovering} are reported. In section \ref{sec:results}, we demonstrate this methodology by considering relatively large Galerkin models of two-dimensional lid driven cavity flow at a Reynolds number $Re = 2 \times 10^4$, where dynamics is chaotic \citep{auteri2002numerical}. We first focus on modal decomposition of the flow and then move to energy analysis and sparsification. Conclusions are offered in section \ref{sec:conclusions}.
 
\section{Methodology}\label{sec:methodology}

\subsection{Reduced Order Modelling}
We consider a space of square integrable velocity vector fields defined over a spatial domain $\Omega$, endowed by the standard inner product
\begin{equation}
    (\mathbf{u},\mathbf{v}) \coloneqq \int_{\Omega} \mathbf{u}\cdot\mathbf{v} \mathrm{d}\Omega, 
\end{equation}
where $\mathbf{u}, \mathbf{v}$ are two elements of such space. The resulting $\mathcal{L}^2(\Omega)$ norm is denoted as \textcolor{black}{$||\mathbf{u}|| = \sqrt{(\mathbf{u}, \mathbf{u})}$}. Using the time averaged velocity field $\bar{\mathbf{u}}(\mathbf{x})$ as a base flow, and denoting by $\mathbf{u}^\prime(t, \mathbf{x})$ the velocity fluctuation $\mathbf{u}(t, \mathbf{x}) - \bar{\mathbf{u}}(\mathbf{x})$, an $N$-dimensional expansion expressed by the ansatz
\begin{equation}\label{eq:modal_expansion}
	\mathbf{u}(t,\mathbf{x}) = \bar{\mathbf{u}}(\mathbf{x}) + \mathbf{u}^\prime(t, \mathbf{x}) = \bar{\mathbf{u}}(\mathbf{x}) + \sum_{i = 1}^{N} a_i(t){\boldsymbol \phi}_i(\mathbf{x}),
\end{equation}
is introduced to describe the space-time velocity field, where $a_i(t)$ and $\boldsymbol \phi_i(\mathbf{x})$, $i=1,\ldots N$ are the temporal and global spatial modes, respectively, with $\|{\boldsymbol \phi}_i(\mathbf{x})\| = 1$. These modes may be computed a posteriori from numerical or experimental data or a priori from a characteristic operator of the system \citep{taira2017modal} or from completeness considerations \citep{NoackEckelmann}. Reduced order models are then derived by projecting the governing equations onto the subspace defined by the modes \citep{rowley2017model}. Restricting our analysis to configurations where the boundaries are either no-slip walls or periodic, this procedure results in an autonomous system of coupled nonlinear ordinary differential equations (ODEs)
\begin{equation}\label{eq:galerkin}
	\sum_{j=1}^N \mathsf{M}_{ij} \dot{a}_j(t) = \tilde{\mathsf{C}}_i + \sum_{j=1}^N \tilde{\mathsf{L}}_{ij}a_{j}(t) + \sum_{j=1}^N \sum_{k=1}^N \tilde{\mathsf{Q}}_{ijk}a_{j}(t)a_{k}(t), \quad \quad i=1\ldots,N,
\end{equation}
defining the temporal evolution of the coefficients $a_{i}(t)$. Here, we only report the definitions of the quadratic coefficients
\begin{equation}\label{eq:definition_qijk}
\tilde{\mathsf{Q}}_{ijk} = (\mathbf{ \boldsymbol{\phi}}_i, \mathbf{\boldsymbol{\phi}}_j \cdot \nabla \mathbf{\boldsymbol{\phi}}_k),
\end{equation}
while expressions for the tensors $\tilde{\mathsfbi{C}}$ and $\tilde{\mathsfbi{L}}$ can be found in \citet{noack2011reduced}. The matrix $\mathsfbi{M}$, with entries $\mathsf{M}_{ij} = (\boldsymbol{\phi}_i,\boldsymbol{\phi}_j)$, takes into account the fact that the spatial modes may not be orthogonal and is introduced here for generality. 

If the $N$ modes span collectively an $N$-dimensional subspace, $\mathsf{M}_{ij}$ is invertible and the system (\ref{eq:galerkin}) can be rearranged as 
\begin{equation}\label{eq:galerkin_2}
	 \dot{a}_i(t) = \mathsf{C}_i + \sum_{j=1}^N \mathsf{L}_{ij}a_{j}(t) + \sum_{j=1}^N \sum_{k=1}^N \mathsf{Q}_{ijk}a_{j}(t)a_{k}(t) \quad \quad i=1\ldots,N ,
\end{equation}
with
\begin{equation}\label{eq:inversion}
     \mathsf{C}_i = \sum_{q=1}^{N}\mathsf{M}^{-1}_{iq}\tilde{\mathsf{C}}_q , \quad \quad \mathsf{L}_{ij} = \sum_{q=1}^{N} \mathsf{M}^{-1}_{iq}\tilde{\mathsf{L}}_{qj} \quad \mathrm{and} \quad \mathsf{Q}_{ijk} = \sum_{q=1}^{N} \mathsf{M}^{-1}_{iq} \tilde{\mathsf{Q}}_{qjk}.
\end{equation}
As observed by \citet{rempfer1994dynamics}, the infinite dimensional matrix $\mathsf{M}_{ij}$ should be first inverted and then truncated to maintain a good prediction accuracy. For the cases discussed in this paper, we have not followed this procedure as we observed that the matrix $\mathsf{M}_{ij}$ has a strong diagonal structure. Hence, the error performed by truncating it to size $(N, N)$ and then inverting it can be reasonably assumed to be small. 

Since the spatial modes satisfy automatically the boundary conditions, the expansion (\ref{eq:modal_expansion}) provides a suitable foundation to examine interactions between coherent structures in complex geometries. Here, we follow established approaches \citep{rempfer1994evolution}) and analyse such interactions by introducing the modal energies $e_i(t) = \frac{1}{2} a_i(t)a_i(t)$, $i=1\ldots,N$. The instantaneous rate of change is given by 
\begin{equation}\label{eq:energy_POD}
    \dot{e}_i(t) = \mathsf{C}_i a_i(t) + \sum_{j = 1}^{N}\mathsf{L}_{ij} a_{i}(t) a_{j}(t) + \sum_{j = 1}^{N}\sum_{k = 1}^{N} \mathsf{Q}_{ijk}a_{i}(t)a_{j}(t)a_{k}(t), \quad \quad i=1,\ldots,N,
\end{equation}
obtained by multiplying (\ref{eq:galerkin_2}) by $a_i(t)$. Note that, in a general case where the modes do not form an orthonormal set, the domain integral of the kinetic energy of velocity fluctuations is given by
\begin{equation}\label{eq:kin_energy}
    E(t) = \frac{1}{2}\int_{\Omega} \mathbf{u}^{\prime}(t,\mathbf{x})^2  d\Omega = \frac{1}{2} \sum_{i=1}^{N}\sum_{j=1}^{N} \mathsf{M}_{ij}a_{i}(t)a_{j}(t)
\end{equation}
and not by a straightforward sum of the terms $e_i(t)$. The right hand side of equation (\ref{eq:energy_POD}) is composed of three terms describing energy transfers between the hierarchy of modes. The first two describe variations of energy due to production/dissipation arising from interactions with the mean flow and from viscous effects \citep{noack2011reduced}. The third term defines variations of energy arising from inviscid nonlinear interactions between triads of modes. Following \citet{rempfer1994dynamics}, these are defined in a time averaged sense by the quadratic interaction tensor $\mathsfbi{N}$ with entries
\begin{equation}\label{eq:non_lin_int}
    \mathsf{N}_{ijk} = \mathsf{Q}_{ijk} \overline{a_i a_j a_k},
\end{equation}
where the overbar denotes temporal averaging. The study of this term is the principal focus of the current analysis. 

Spatial modes obtained from classical decompositions have generally global support over the domain (see e.g. \citet{taira2017modal}). The result is that the evolution equations (\ref{eq:galerkin_2}) are not strictly sparse in the sense employed by \citet{brunton2016discovering}. In fact, unless particular symmetries apply, the tensor $\mathsfbi{Q}$ is generally dense, \textit{i.e.} most of its entries are different from zero and the right hand side of \eqref{eq:galerkin_2} contains all monomial terms in the modal amplitudes $a_i(t)$ up to order two.
%
However, as anticipated in the introduction, in turbulent realisations of the Navier-Stokes equations only a subset of all triadic interactions contributes to a significant degree to the overall energy budget \citep{couplet2003intermodal, rempfer1994evolution}. In this sense, sparsity is a primarily an \textit{a posteriori} feature of solutions, i.e. a feature of the quadratic interaction tensor $\mathsfbi{N}$. 

The approach developed in this work starts from this fundamental observation and aims to generate a sparse Galerkin model, defined by a sparse coefficient tensor $\mathsfbi{Q}^{s}$ that is a good approximation of the original dynamical system in the sense that the mismatch between the transfer tensors $\mathsfbi{N}^s$ and the original $\mathsfbi{N}$ obtained from the definitions (\ref{eq:definition_qijk}, \ref{eq:inversion}) is as small as possible across the hierarchy of modes.

\subsection{Sparse regression}
To construct a sparse Galerkin system, we use a procedure akin to that utilised in previous work for calibrating Galerkin models from data (\cite{perret2006polynomial, cordier2010calibration, Xie:2018dv}) and more recently for the identification of sparse dynamical systems \citep{brunton2016discovering}. In the first step, we assume that $N_t$ snapshots of the velocity field are available from simulation and arrange samples of the temporal coefficients $a_i(t_j)$, $i=1,\ldots,N$ and $j=1,\ldots,N_t$, into the data matrix $\mathsfbi{A} \in \Re^{N_t \times N}$, with entries $\mathsf{A}_{ij} = a_i(t_j)$.
Similarly, we construct the modal acceleration matrix $\dot{\mathsfbi{A}} \in \Re^{N_t \times N}$, containing the time derivative of the temporal coefficients obtained by projecting the modes $\boldsymbol{\phi}_i(\mathbf{x})$ on snapshots of the Eulerian acceleration field $\partial_t \mathbf{u}(t_j, \mathbf{x})$ and correcting such projections with $\mathsfbi{M}$ when modes are not orthogonal (see also \citet{rempfer1994evolution}). We then exploit the polynomial structure of the Galerkin system (\ref{eq:galerkin_2}) to construct the database matrix ${\boldsymbol \Theta}(\mathsfbi{A}) \in \Re^{N_t \times q}$
\begin{equation}\label{eq:database}
  {\boldsymbol \Theta}(\mathsfbi{A}) = \begin{pmatrix}
	1 & a_1^1 & a_2^1 & \hdots & a_N^{1}& a_1^1 a_1^1 & \hdots & a_N^{1} a_N^{1}\\
 	\vdots        & \vdots        & \vdots&   & \vdots  & \vdots      &        & \vdots\\
    1 &  a_1^{N_t} &  a_2^{N_t} & \hdots & a_N^{N_t}  & a_1^{N_t} a_1^{N_t} & \hdots & a_N^{N_t} a_N^{N_t}
	\end{pmatrix},
\end{equation}
called nonlinear feature library in \citet{brunton2016discovering}, where $q = (N+1) + N(N+1)/2$ is the total number of features, the sum of constant, linear and quadratic interactions. The number of quadratic coefficients is only $N(N+1)/2$ because the interaction between mode $i$ and $j$ is considered only once in (\ref{eq:database}). As discussed later on in the paper, this avoids columns of ${\boldsymbol \Theta}(\mathsfbi{A})$ becoming linearly dependent, which would in turn result in numerical stability issues in the solution regression problem (see e.g. \citet{perret2006polynomial} and \citet{cordier2010calibration}).

Arranging the projection coefficients tensors $\mathsfbi{C}$, $\mathsfbi{L}$ and $\mathsfbi{Q}$ associated to the $i$-th mode into a coefficient vector ${\boldsymbol \beta}_i\in \Re^{q}$, the Galerkin system (\ref{eq:galerkin}) can be equivalently expressed as
\begin{equation}\label{eq:system-for-each-mode}
    \dot{\mathsfbi{A}}_i = {\boldsymbol \Theta}(\mathsfbi{A}){\boldsymbol \beta}_i, \quad i = 1, \ldots N,
\end{equation}
where $\dot{\mathsfbi{A}}_i$ is the $i-$th column of the modal acceleration matrix. The key idea is that if some nonlinear interactions are more important than others, then the corresponding entries of the coefficient vector ${\boldsymbol \beta}_i$ can be shrunk to zero with minor effects on the predictive ability of the resulting model. The challenge is to find a systematic method to identify the dominant interactions and prune unnecessary coefficients whilst calibrating the remaining model coefficients such as to preserve the overall energy budget. Here, we adopt an established sparsity-promoting regression technique known as LASSO regression (Least Absolute Shrinkage Selection Operator, see \citet{tibshirani1996regression}). In short, it leads to a set of $N$ optimisation problems of the form
\begin{equation}\label{eq:lasso}
\displaystyle \underset{{\boldsymbol \beta}_i}{\mathrm{min}}  ||{\boldsymbol \Theta}(\mathsfbi{A}) {\boldsymbol \beta}_i - \dot{\mathsfbi{A}}_i||_2^2 +  \gamma_i||{\boldsymbol \beta}_i||_1, \quad \quad i=1,...,N,
\end{equation}
one for each mode, where $\|\cdot\|_p$ denotes the $l_p$ norm of a vector.
The first term in the objective function in (\ref{eq:lasso}) produces calibrated models that have minimum prediction error on the modal acceleration (see discussion in \citet{cordier2010calibration} and \citet{couplet2005calibrated}). The second term penalises large model coefficients, regularises the regression and encourages sparsity in the solution by shrinking exactly to zero coefficients in ${\boldsymbol \beta}_i$ corresponding to \textcolor{black}{columns of} ${\boldsymbol \Theta}(\mathsfbi{A})$ with little dynamical influence. Ideally, to prune unnecessary coefficients, a penalisation term proportional to the cardinality of ${\boldsymbol \beta}_i$, $\mathrm{card}({\boldsymbol \beta}_i)$, would formally be more correct \citep{jovanovic2014sparsity}. However, the resulting optimisation problem would be computationally intractable even for Galerkin models of modest dimensions. In fact, this penalisation is usually relaxed to the computationally tractable $l_1$ term \citep{ramirez2013l1}. Regardless, the optimisation problems (\ref{eq:lasso}) are convex and thus have an unique solution. In addition, the approach lends naturally to parallelisation, since the optimisation problems can be solved independently for each mode. In initial stages of the research, we have found approaches based on sequential thresholded least-squares (\cite{brunton2016discovering,zhang2019convergence, loiseau2018constrained}) to be not sufficiently robust. Hence, solutions of (\ref{eq:lasso}) have been computed using the \texttt{sklearn} (\cite{scikit-learn}) library, which implements a sub-gradient descent algorithm to manage the non differentiability of the $l_1$ norm.

The \textcolor{black}{weights} $\gamma_i$ in equation (\ref{eq:lasso}) are arbitrary \textcolor{black}{and can be tuned to} trade prediction ability (when they are small) for sparsity (when they are large).
To formalise these concepts, we introduce the global reconstruction error $\epsilon$
\begin{equation}\label{eq:epsilon}
    \epsilon =  \displaystyle\sum_{i=1}^{N} \frac{\displaystyle ||{\boldsymbol \Theta}(\mathsfbi{A}){\boldsymbol \beta}_i - \dot{\mathsfbi{A}}_i||^2_2}{\displaystyle ||\dot{\mathsfbi{A}}_i||^2_2}
\end{equation}
and the \textcolor{black}{system density} $\rho$
\begin{equation}\label{eq:rho}
    \rho = \frac{1}{N  q} \displaystyle \sum_{i=1}^{N} \mathrm{card}({\boldsymbol \beta}_i).
\end{equation}
In equation (\ref{eq:epsilon}), the absolute reconstruction error $||{\boldsymbol \Theta}(\mathsfbi{A}){\boldsymbol \beta}_i - \dot{\mathsfbi{A}_i}||^2_2$ is normalised with the mean squared acceleration $||\dot{\mathsfbi{A}_i}||^2_2$ to balance the global reconstruction error across the hierarchy, which would be otherwise dominated by the most energetic modes. On the other hand, the density $\rho$ ranges from $0$, when all interactions have been pruned, to $1$, for a fully connected model. Note that for large models, the density is dominated by the quadratic tensor $\mathsfbi{Q}$. A one-parameter family of models can be generated by varying the regularisation weight\textcolor{black}{s} $\gamma_i$, producing a Pareto front \citep{Schmidt:2009dt} on the $\rho$--$\epsilon$ plane. Since only a subset of triadic interactions is relevant, the expectation is that a sweet spot appears on this curve, defining \textcolor{black}{`optimal' penalisation coefficients} $\gamma_i$. 
\textcolor{black}{It is important to observe that the penalisation coefficient $\gamma_i$} can be chosen independently for each index $i$, \textcolor{black}{implying that reconstruction error and sparsity can be modulated arbitrarily across the spectrum of modes.}
In our analysis we consider two different modulation strategies. In strategy S1, the weight is constant for all modes, $\gamma_i = \gamma$. This strategy sparsifies more aggressively the equations of motion of low-energy modes, because the $l_1$ penalisation term has a higher importance than the $l_2$ component. \textcolor{black}{In this work we observed that the lowest global reconstruction error is obtained when $\gamma_i$ is kept constant across the modes. We} also introduce strategy S2, where the weight is normalised with respect to the mean squared modal acceleration as $\gamma_i = ||\dot{\mathsfbi{A}}_i||_2^2 \gamma$. This is equivalent to solving problem \eqref{eq:lasso} using the relative error in \eqref{eq:epsilon} as least-squares component of the objective function. This strategy results in a more balanced sparsification across the hierarchy of modes and avoids earlier truncation, i.e. when all coefficients \textcolor{black}{of a high-index mode} are set to zero. \textcolor{black}{Other strategies can be, of course, devised. Here, we mention, for instance, the possibility to tune the penalisation coefficients to obtain a uniform sparsification across the spectrum or to obtain a uniform relative reconstruction error. Analysing these strategies is an interesting avenue for future work.}

\textcolor{black}{One potential modification of this approach is that discussed in \citet{loiseau2018constrained}, namely to enforce that the nonlinear term in the sparsified Galerkin model conserves energy exactly (see e.g. \citet{balajewicz2013low} and \citet{noack2008finite} for a formal definition). In practice, this can be achieved by introducing a set of constraints on the coefficients vectors ${\boldsymbol \beta}_i$. The constraints, however, couple together the regression problems of all modes, resulting in one optimisation problem of larger dimension. As we will demonstrate later in the paper, the energy conservation error of models obtained from the unconstrained approach is small in relative terms. This occurs because the temporal coefficients in $\mathsfbi{A}$ are originally obtained from an energy conserving nonlinearity, and the regression ``discovers'' this property from data. Hence, throughout this paper we always solved problems (\ref{eq:lasso}) independently, without additional constraints.}

\section{Results}\label{sec:results}
\textcolor{black}{We now apply this methodology to two-dimensional unsteady flow in a lid-driven square cavity. This is an established test case for the development and validation of model order reduction techniques \citep{cazemier1998proper, terragni, balajewicz2013low, arbabi2017study, fick2017reduced}, and we thus consider it here as an exemplar to demonstrate the ideas discussed in the introduction.}

The Reynolds number is defined as $Re = LU/\nu$ where $L$, $U$ are the cavity dimension and the lid velocity, respectively, while $\nu$ is the kinematic viscosity. \textcolor{black}{These quantities are used to make the equations of motion nondimensional. We purposefully investigate a regime at $Re = 2 \times 10^4$, where the flow evolves in a chaotic manner (\citet{auteri2002numerical, PENG2003337}, see also the animation of the vorticity field in the supplementary material). The chaotic nature of the problem ensures that the frequency spectrum of velocity fluctuations is continuous and energy transfers are scattered in modal space, rather than being highly organised as for periodic flows \citep{noack2011reduced}}. The domain is defined by \textcolor{black}{the nondimensional Cartesian coordinates $\mathbf{x} = (x, y)$ and the velocity vector $\mathbf{u}(t, \mathbf{x})$ is defined by the components $u(t, \mathbf{x})$ and $v(t, \mathbf{x}))$}. For visualisation purposes, we introduce the out-of-plane vorticity $\omega = \partial v/\partial x - \partial u/\partial y$.
\begin{figure}
  \centerline{\includegraphics[width=0.98\textwidth]{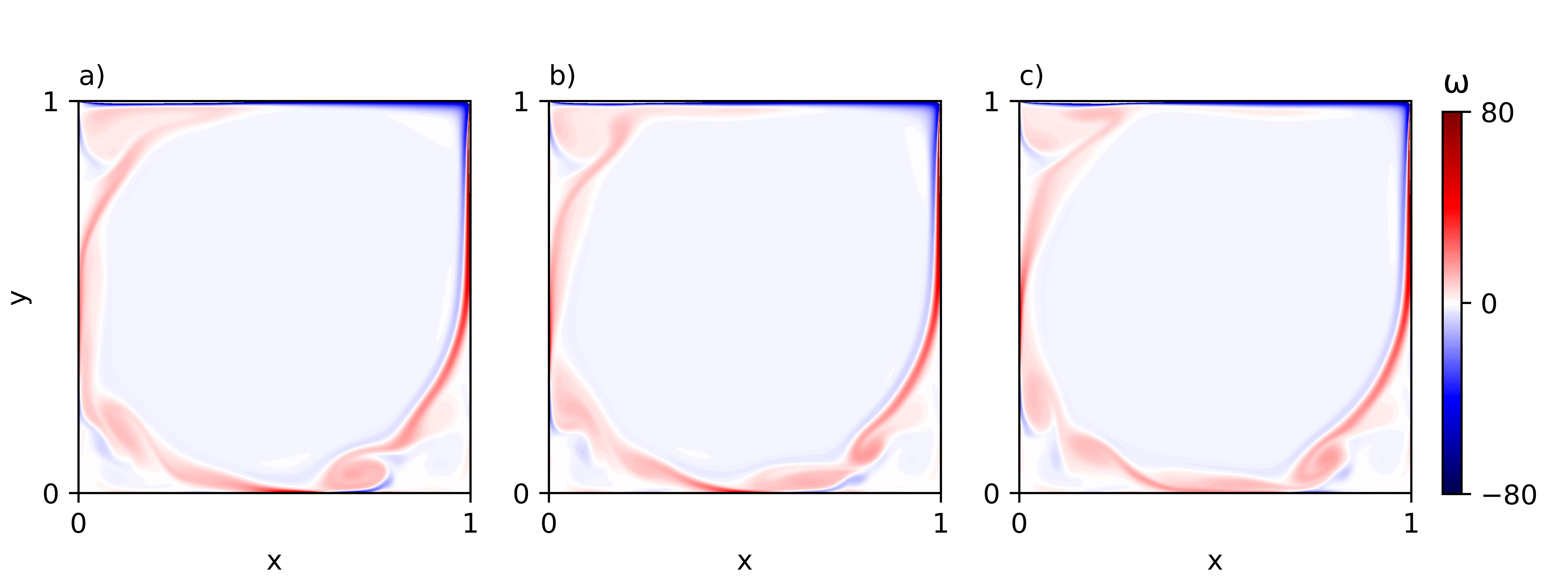}}
  \caption{Vorticity field $\omega$ of three different snapshots separated by one non-dimensional time unit, increasing from left to right. \textcolor{black}{An animation of the vorticity field, illustrating the chaotic nature of the dynamics and the dominant flow features is available in the supplementary material.}} \label{fig:validation}
\end{figure}

Numerical simulations were performed in \texttt{OpenFOAM} with the incompressible flow solver \texttt{icofoam}. The convective and viscous terms are spatially discretised with a second-order finite-volume technique and the temporal term with a semi-implicit Crank-Nicholson scheme. \textcolor{black}{Special treatments of the singularities developing at the top corners due to the discontinuity in the velocity boundary conditions \citep{BOTELLA1998421} were not deemed necessary}.
\textcolor{black}{A grid independence study was initially performed, examining average and unsteady flow quantities on increasingly finer meshes. The final mesh is composed of $300 \times 300$ cells, with refinement at the four cavity boundaries. This mesh is sufficiently fine to resolve the unsteady high shear regions bounding the main vortex, the high vorticity filaments characteristic of two-dimensional turbulence as well as the spatial structure of the lowest energy modes utilised for the projection. Similar grid resolutions have been used by \cite{cazemier1998proper} at similar Reynolds numbers}. 


Three snapshots of the vorticity field obtained \textcolor{black}{from these simulations} are shown in figure \ref{fig:validation} \textcolor{black}{(see also the animation in the supplementary material). Most of the dynamically interesting features in this regime originate at the bottom-right corner of the cavity. Specifically, the secondary vortex in the recirculation zone is shed erratically, producing wave-like disturbances advected along the shear layer bounding the primary vortex}. The characteristic non-dimensional frequency of this \textcolor{black}{wave-like} motion is \textcolor{black}{$f = 0.7$}. From \textcolor{black}{simulation}, we extract $N_t = 1500$ velocity snapshots using a nondimensional sampling period $\Delta t = 0.1$. These settings are sufficient to adequately time-resolve the fast scales as well as to include many shedding events at the bottom right corner, making the regression problems (\ref{eq:lasso}) statistically reliable. 

\subsection{Modal Decomposition}\label{sec:modal_decomposition}
First, we consider models generated using POD modes. POD produces economic reduced order models, but has the well-known shortcoming of mixing together fluid motions at different temporal/spatial scales \citep{mendez2018multi}. 
Second, we consider models generated from modes oscillating at a single frequency obtained from a procedure that is equivalent to a Discrete Fourier Transform (DFT) of the velocity snapshots. \textcolor{black}{For practical convenience, we obtain the two distinct sets of modes using the same computational technique, based on the approach proposed by \citet{sieber2016spectral} which only operates on the temporal correlation matrix}. Briefly, the method considers the temporal correlation matrix $\mathsfbi{R} \in \mathbb{R}^{N_t, N_t}$, with entries
\begin{equation}\label{eq:corr}
    \mathsf{R}_{ij} = \frac{1}{N_t}\left( \mathbf{u}^\prime(t_i, \mathbf{x}), \mathbf{u}^\prime(t_j, \mathbf{x}) \right),
\end{equation}
and then defines a filtered correlation matrix $\mathsfbi{S}$, with elements
\begin{equation}\label{eq:filtering}
    \mathsf{S}_{ij} = \sum_{k = -N_f}^{k = N_f} g_k \mathsf{R}_{i+k,j+k}
\end{equation}
given by the application of the filter coefficient vector $\mathbf{g}$ along the diagonals of the correlation matrix. An ordered set of temporal coefficients \mbox{$\mathbf{a}_i = [a_i(t_1), \ldots, a_i(t_{N_t})]$} and associated mode energies $\lambda_i$ is then obtained from the eigendecomposition of $\mathsfbi{S}$,
\begin{equation}
    \mathsfbi{S} \mathbf{a}_i = \lambda_i \mathbf{a}_i,
\end{equation}
so that $\lambda_i\delta_{ij} = \mathbf{a}_i^\top\cdot\mathbf{a}_j$. As discussed in \citet{sieber2016spectral}, when the filter is extended over the entire dataset and in the limit of number of samples tending to infinity, the filtered correlation matrix converges to a Toeplitz, circulant matrix. Then, its eigenvalues trace the power spectral density of the underlying data set. On the other hand, the eigenvectors $\mathbf{a}_i$ corresponds to the Fourier basis. This procedure generates conjugate pairs of modal structures with same energy oscillating at a single frequency. These can be viewed as a set of modal oscillators exhibiting periodic fluctuations \citep{taira2017modal} and tracing fluid motion at on a two-dimensional subspace. In practice, for a finite-length dataset, we filter the temporal correlation matrix assuming periodicity using a box-car filter, as suggested in \citet{sieber2016spectral}. Hereafter, we will refer to the modal structures identified by this procedure as DFT modes.

One important consideration is that, unlike Dynamic Mode Decomposition (see \citet{Rowley:2009ez, schmid2010dynamic, Chen:2012jh}), DFT lacks the ability to discern and identify dominant frequency components. Instead, a number of modes equal to the number of snapshots utilised is produced, oscillating in conjugate pairs at specific frequencies determined by the sampling period $\Delta t $ and observation time $T$ \citep{mendez2018multi}. This property, \textit{picket fencing}, results in frequencies that are integer multiples of the fundamental frequency $f_1 = T^{-1}$, up to the Nyquist component $f_{Nyq} = (2\Delta t)^{-1}$. In addition, unlike for POD, as the length of the dataset is increased, the number of energy-relevant modes increases and low-frequency modes with little dynamical importance appear. The approach we use here is to divide the dataset into five partition of thirty time units, covering an average of 20 cycles of the dominant oscillatory component, and providing sufficient frequency resolution to distinguish small scale spectral features. In addition, two possible ways of sorting pairs of modal structures are possible, i.e. by energy content (using the eigenvalues $\lambda_i$) or by frequency. Models obtained with the two sorting schemes will be referred to as $\mathrm{DFT}_e$ and $\mathrm{DFT}_f$, respectively.

\begin{figure}
  \centerline{\includegraphics[width=0.85\textwidth]{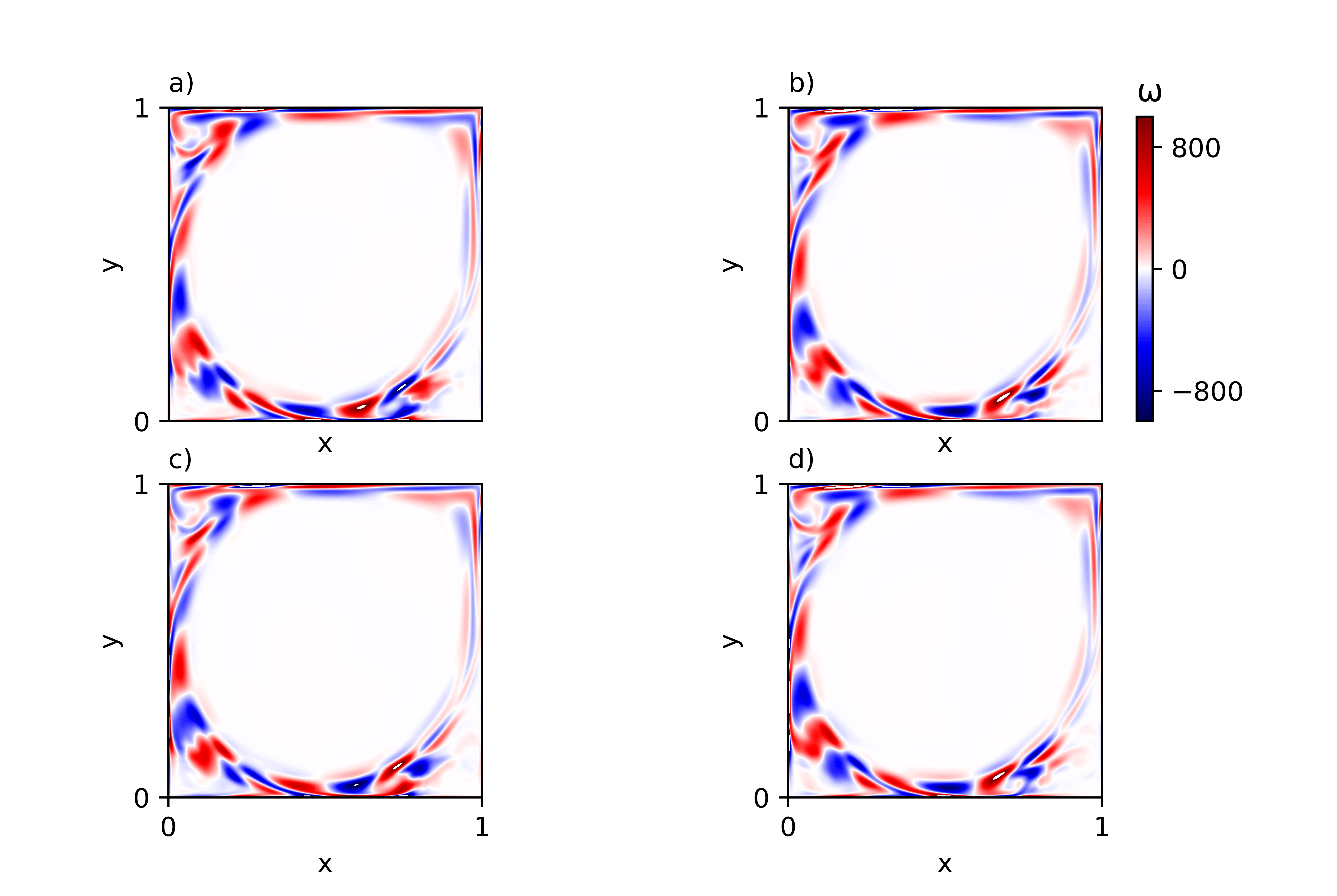}}
  \caption{Vorticity field of the most energetic pair of DFT modes, panels (a) and (b), and of the first two POD modes, panels (c) and (d).}
  \label{fig:spatial_scales}
\end{figure}

We now focus on the characteristics of the modal structures obtained by these two methods. We denote the normalised cumulative sum of the eigenvalues $\lambda_i$ of the (filtered) correlation matrix as
\begin{equation}\label{eq:resolved_energy}
    \displaystyle e(n) = {\displaystyle\sum_{i=1}^n \lambda_i }/{\displaystyle\sum_{i=1}^{N_t} \lambda_i},
\end{equation}
describing the fraction of the fluctuation kinetic energy captured by the first $n$ elements of the expansion (\ref{eq:modal_expansion}). 
\begin{table}
  \begin{center}
\def~{\hphantom{0}}           
  \begin{tabular}{lllllllllllll}
       $n$  & 1 & 5 & 10 & 15 & 20 & 26 & 50 & 75 &80  & 95 & 100 & 300 \\[3pt]
       \hline
       POD   & 0.26 & 0.74 & 0.85 & 0.89 & 0.9 & 0.95 & 0.98& 0.99 & 0.995 & 0.998 & 0.999 & -     \\
       $\mathrm{DFT}_e$  & 0.17 & 0.49 & 0.62 & 0.74 & 0.8 & 0.88  & & 0.92  & 0.95 & 0.97 & 0.98 & 1 \\
       $\mathrm{DFT}_f$ & 0.02 & 0.07 & 0.15 & 0.39 & 0.43 & 0.47  & 0.91&  & 0.96 & 0.97 & 0.98 & 1
  \end{tabular}
  \caption{Normalised cumulative energy distribution $e(n)$ for POD and DFT modes, where the latter are sorted by energy content ($\mathrm{DFT}_e$) or by frequency ($\mathrm{DFT}_f$).}
  \label{tab:kd}
  \end{center}
\end{table}
\begin{figure}
  \centerline{\includegraphics[width=1\textwidth]{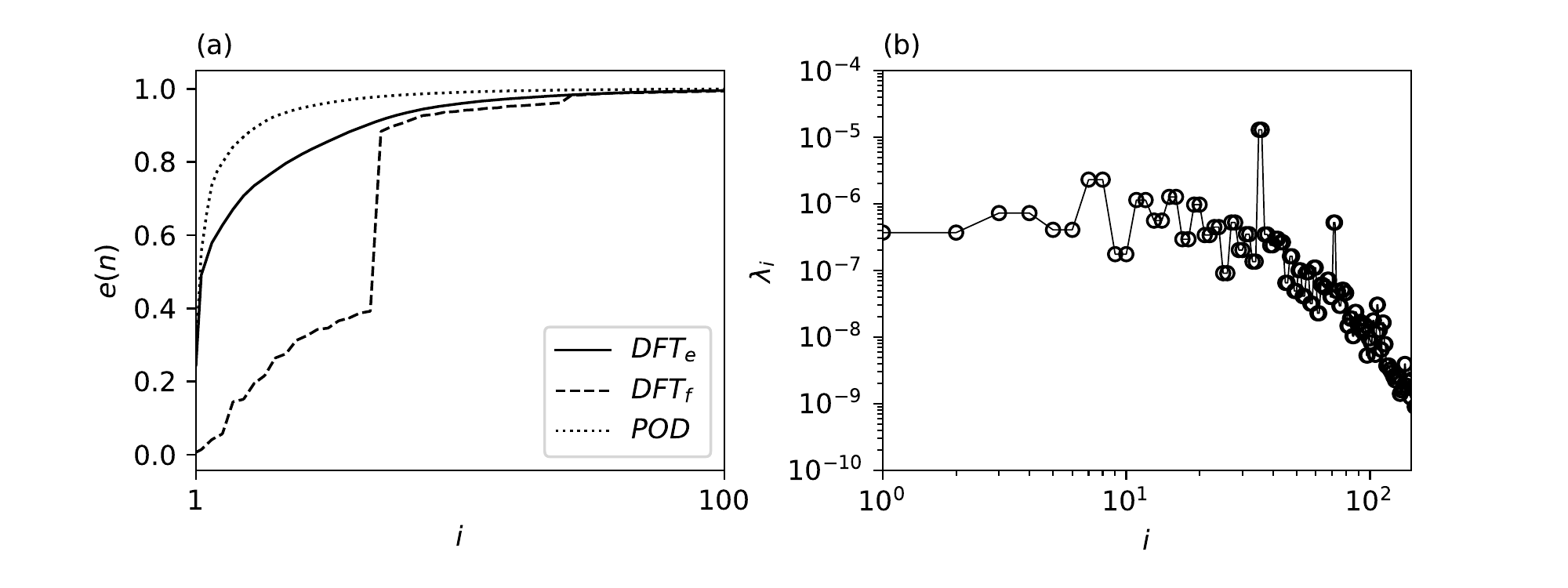}}
  \caption{Panel (a): cumulative sum of the first $100$ eigenvalues of $\mathsf{S}_{ij}$ for the three decompositions considered. Panel (b): distribution of the modal energies of $\mathrm{DFT}$ modes sorted by frequency.}
  \label{fig:eigenvalues}
\end{figure}
This quantity is shown in figure \ref{fig:eigenvalues}-(a) for the POD and for the two possible DFT sorting schemes.  As expected, a larger energy is captured by the POD basis. For the DFT decomposition, the energy-based sorting is more efficient at data compression, although the difference vanishes for large $n$, since for low energy modes the two sorting schemes are equivalent. The modal energies associated to the $\mathrm{DFT}_f$ modes are shown in figure \ref{fig:eigenvalues}-(b) as a function of the modal index $i$. The distribution is characterised by a continuous component, with modal energy decaying with frequency, and a discrete component, with a fundamental peak for the pair of modes (31, 32) and its first few harmonics. The peak, at a non-dimensional frequency $f =  0.7$, is physically originated from the high-energy structures transported along the shear layer by the rotation of the main vortical structure.  
This can be observed in panels (a) and (b) of figure \ref{fig:spatial_scales}, showing the vorticity field $\omega$ of the DFT mode pair (31, 32). 
This pair of modes describes a vorticity perturbation having the form of a wave travelling along the edge of the main vortex. Hence, the spatial structure of the two modes is shifted in the direction of the shear layer by half wave. \textcolor{black}{Travelling-wave structures in cavity flows have already been observed in simulation by \citet{poliashenko1995direct,auteri2002numerical} and characterised by global stability analysis and Koopman analysis by \citet{boppana} and \citet{arbabi2017study}, respectively}. The two leading POD modes, reported in panels (c) and (d) of figure \ref{fig:spatial_scales}, have the same energy and \textcolor{black}{capture the same travelling-wave pattern described by the leading DFT mode pair}.



\subsection{Energy Analysis}
To provide a more robust foundation to understand the sparsification results reported in sections  \ref{sec:sparsification_POD} and \ref{sec:sparsification_DFT}, we first focus on the analysis of the average energy interactions.
\begin{figure}
  \centerline{\includegraphics[width=1.1\textwidth]{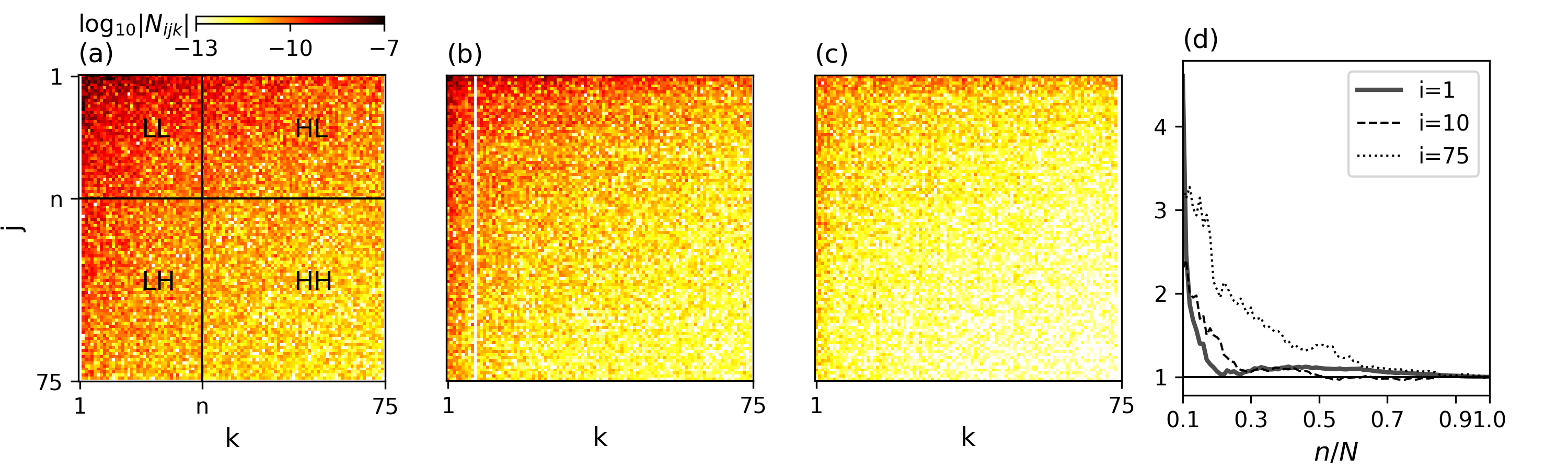}}
  \caption{Magnitude of the average interaction tensor coefficients $\mathsf{N}_{ijk}$ for three POD modes across the spectrum, $i = 1, 10$ and 75 in panel (a), (b) and (c) respectively, for a model resolving 99\% of the fluctuation kinetic energy. Panel (d) shows the coefficient $\chi_i(n)$ as a function of the normalised cutoff $n$ for the same three modes.}
   \label{fig:N_ijk_POD}
\end{figure}
The structure of the interaction tensor $\mathsfbi{N}$ for a large POD-based model with $N=75$,  reconstructing more than $99\%$ of the fluctuation kinetic energy, is reported in figure \ref{fig:N_ijk_POD}, showing the magnitude of the interactions for three slices for $i=1, 10$ and $75$, in panels (a), (b) and (c), respectively. All entries of the tensor $\mathsfbi{N}$ are generally nonzero, although the strength of the interactions varies across several orders of magnitude. This is a combined result of the projection coefficients tensor $\mathsfbi{Q}$ \textcolor{black}{(shown later)}, whose entries are typically non zero, and of the complex spectral structure of the temporal coefficients $a_{i}(t)$.  The most important feature of figure \ref{fig:N_ijk_POD} is that interactions are highly organised and there exists a subset of interactions that are more active. Specifically, for any mode $i$, triadic interactions can be classified as illustrated in panel (a) in four different categories by introducing a cutoff modal index $n$. The subset of interactions denoted as $\mathrm{LL}$ corresponds to nonlinear energy transfer involving pairs of low index modes, $\mathrm{HL}$ and $\mathrm{LH}$ denote interactions involving high-low/low-high \textcolor{black}{index} modes, while $\mathrm{HH}$ denotes the subset of interactions involving pairs of high \textcolor{black}{index} modes. We observe that the areas corresponding to $\mathrm{LL}$ and $\mathrm{HL}/\mathrm{LH}$ are the most active. If we map low/high modal indices to large/small scales, this result is in agreement with the picture of energy transfer between scales in homogeneous isotropic two-dimensional turbulence \citep{ohkitani1990nonlocality,laval1999nonlocality}, where the large scales interact with the small ones in a non-local fashion. In addition, interactions are not symmetric with respect to a swap of indices $j, k$. This can be quantified by computing the coefficient 
\begin{equation}\label{eq:chi}
    \chi_i(n) = \sum_{j=1}^{n} \sum_{k=1}^{N} \mathsf{N}_{ijk} / \sum_{j=1}^{N} \sum_{k=1}^{n} \mathsf{N}_{ijk},
\end{equation}
representing the relative dynamical importance of the subset of interactions $\mathrm{LL}+\mathrm{HL}$ and $\mathrm{LL}+\mathrm{LH}$. Panel (d) of figure \ref{fig:N_ijk_POD} shows $\chi_i$ for $i = 1,10$ and $75$ as a function of the normalised cutoff $n$.  The interaction subset $\mathrm{HL}$ is up to four times more important than the subset $\mathrm{LH}$. This is a consequence of the asymmetry of the projection coefficients $\mathsf{Q}_{ijk}$, which arise from the fact that the convective transport of structure $\mathbf{\boldsymbol{\phi}}_k(\mathbf{x})$ operated by the structure $\mathbf{\boldsymbol{\phi}}_j(\mathbf{x})$ is more intense when the modal structure $\mathbf{\boldsymbol{\phi}}_j(\mathbf{x})$ describes large-scale flow features.

\begin{figure}
  \centerline{\includegraphics[width=1\textwidth]{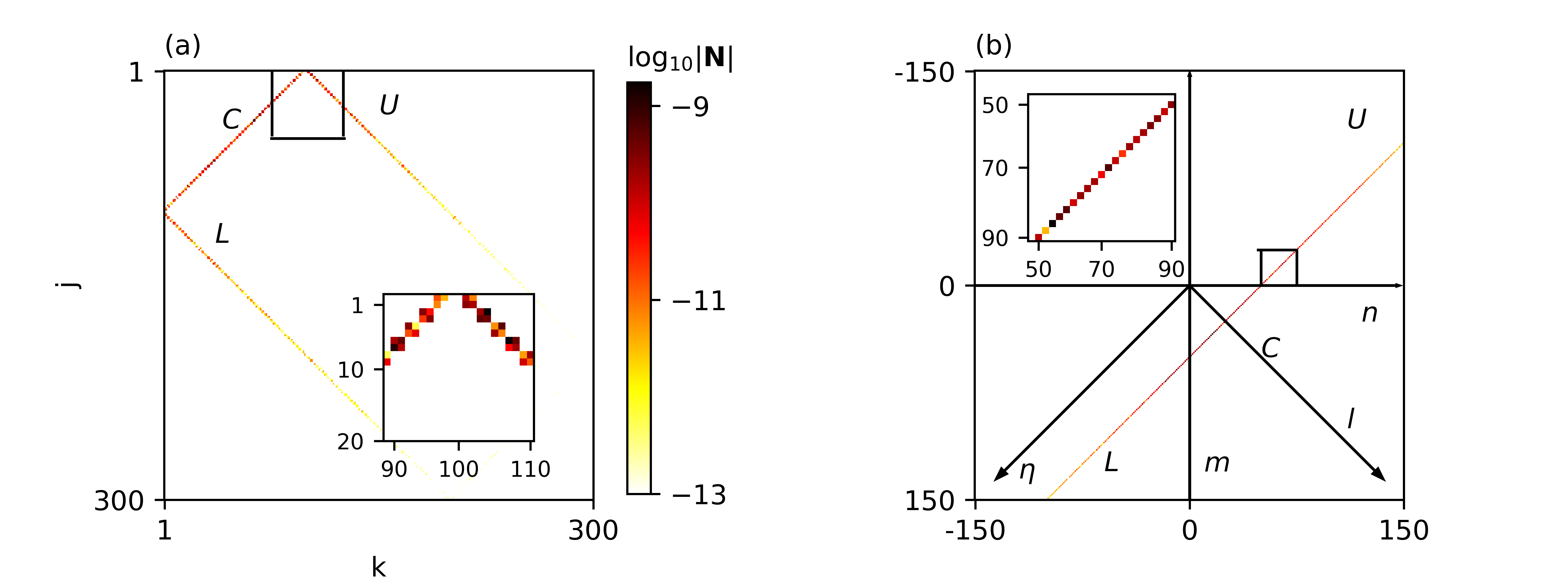}}
  \caption{Panel (a): Magnitude of the average interaction tensor $\mathsf{N}_{ijk}$ for $i=100$, with the three characteristics branches, showing that active interactions come in $2\times 2$ blocks corresponding to matching triads of modes. The small inset focuses on the interactions  of branches C and U. Panel (b): magnitude of the average interaction tensor \eqref{eq:energy_coherent_structures} where the three branches of panel (a) have been unfolded on a larger plane spanned by the coordinates $l$ and $\eta$. The inset shows details of the interactions of the branch U in the plane $\eta-l$.}
  \label{fig:Nijk_sum}
\end{figure}

We now consider energy analysis of a large, full-resolution $\mathrm{DFT}_f$ model constructed from five partitions of thirty time units as discussed in section \ref{sec:modal_decomposition}. The model is composed of all $N=300$ modes, corresponding to $150$ distinct frequencies. We perform modal decomposition and energy analysis on each partition separately, and then average the mean energy transfer rate tensor $\mathsfbi{N}$ over the five partitions. Figure \ref{fig:Nijk_sum}-(a) shows the mean transfer rate distribution for mode $i=100$.
Energy interactions in the DFT model are very sparsely distributed on a thin horseshoe-shaped structure composed of three branches (denoted in the figure as L, C and U) of $2\times 2$ blocks, and all other mean energy transfer rates interactions are identically zero. This pattern results from the joint effect of the oscillatory nature of the temporal coefficients and the quadratic nonlinearity of system (\ref{eq:galerkin}), which can only be satisfied by triads of modes having matching temporal wave numbers.
A less pronounced horseshoe-shaped distribution of the energy interactions has been previously observed in energy analysis of POD-based models of three-dimensional transitional boundary layers \citet{rempfer1994dynamics,rempfer1994evolution}. These authors noticed that low-energy modal structures resembles Fourier modes in the spanwise direction \textcolor{black}{(justified by the spanwise periodic domain)} and thus coefficients $\mathsf{Q}_{ijk}$ and energy interactions are nonzero only for specific triads of modes. In the present case, this pattern is determined exclusively by the temporal coefficients as the tensor $\mathsfbi{Q}$ constructed from projection modes does not posses any \textcolor{black}{sparsity structure} and its coefficients have a similar statistical distribution to that obtained using the POD modes. 
\begin{figure}
  \centerline{\includegraphics[width=1\textwidth]{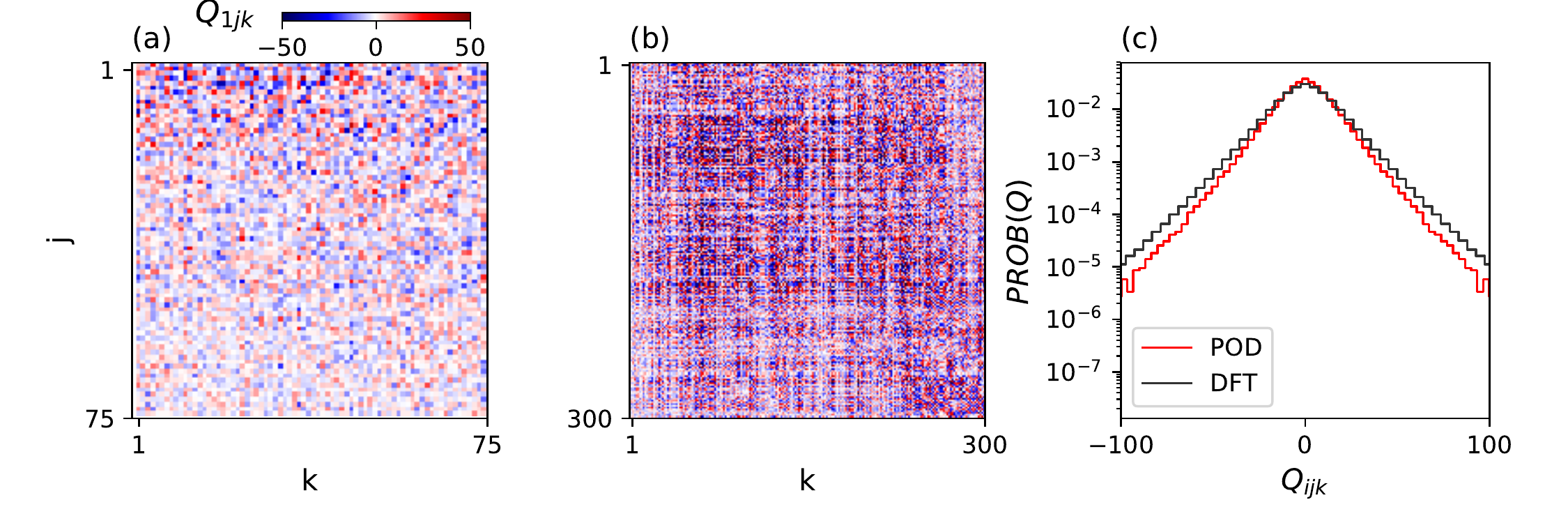}}
  \caption{\textcolor{black}{Maps of $\mathsf{Q}_{1jk}$ for Galerkin models constructed from the POD and the $\mathrm{DFT}_f$ decompositions, in panel (a) and (b), respectively. Panel (c) shows the probability distribution of all quadratic coefficients for these two models.}}
  \label{fig:Q_coeffs}
\end{figure}
\textcolor{black}{This is illustrated in figure \ref{fig:Q_coeffs} showing maps of the first slice of the tensor $\mathsfbi{Q}$ of the largest Galerkin models considered here, constructed from the POD and the DFT decompositions, in panels (a) and (b) respectively. We observe that no underlying structure is present except for the asymmetry already observed in the energy analysis in figure \ref{fig:N_ijk_POD}. This property is confirmed in the probability distribution of the coefficients, shown in panel (c).}

%

\begin{figure}
  \centerline{\includegraphics[width=1\textwidth]{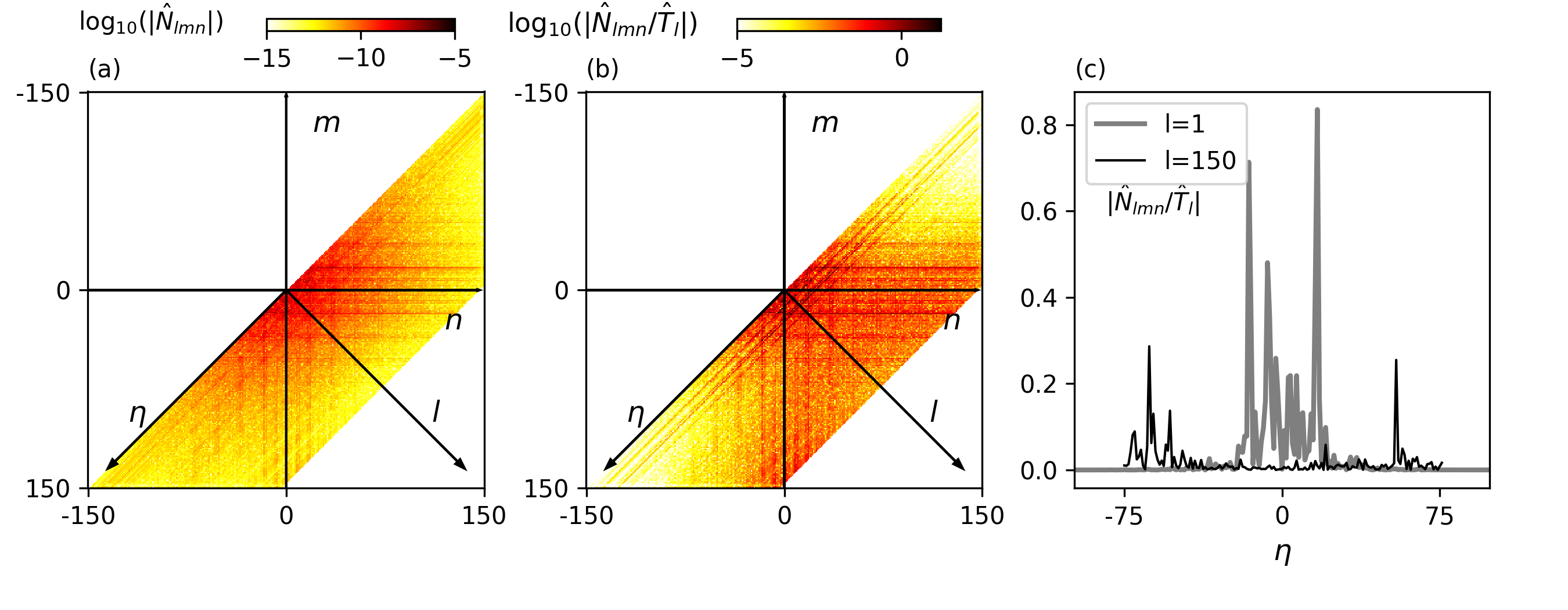}}
  \caption{Panels (a) and (b): absolute and relative strength of the energy interactions between pairs of DFT modes for a model with $N=300$ visualised on the plane $m,n$, with the additional coordinates $l$ and $\eta$. Panels (c): relative energy interactions for the first and last mode pairs.}
  \label{fig:Energy_cuts}
\end{figure}

To facilitate the interpretation of the energy interaction pattern, we follow \citet{rempfer1994evolution} and \citet{arbabi2017study} and define oscillatory modal structures
\begin{equation}
    \mathbf{u}_l(t,\mathbf{x}) =  a_{2l-1}(t){\boldsymbol \phi}_{2l-1}(\mathbf{x}) + a_{2l}(t){\boldsymbol \phi}_{2l}(\mathbf{x}),
\end{equation}
numbered by the index $l$ and tracing fluid motion at a single frequency on a two-dimensional subspace. Their modal energy is
\begin{equation}
    e_l(t) = \frac{1}{2}\left(a^2_{2l-1}(t) + a^2_{2l}(t)\right) + a_{2l-1}(t) a_{2l}(t) ({\boldsymbol \phi}_{2l-1}(\mathbf{x}),{\boldsymbol \phi}_{2l}(\mathbf{x})).
\end{equation}
Numerical experiments show that, for large number of snapshots, pairs of modes  ${\boldsymbol \phi}_{2l-1}(\mathbf{x})$ and ${\boldsymbol \phi}_{2l}(\mathbf{x})$ tend to be orthogonal. Hence, considering the evolution equation for the modal energy $e_l(t) \sim \frac{1}{2}(a^2_{2l-1}(t) + a^2_{2l}(t))$ leads to the condensed triadic interaction tensor $\hat{\mathsfbi{N}}$ of size $(N/2, N/2, N/2)$ with entries
\begin{equation}\label{eq:energy_coherent_structures}
    \hat{\mathsf{N}}_{lmn} = \displaystyle \sum_{i=l}^{l+1} \sum_{j=m}^{m+1} \sum_{k=n}^{n+1}\mathsf{N}_{ijk},
\end{equation}
lumping together the $2\times 2$ blocks of interactions at matching triads of figure \ref{fig:Nijk_sum}-(a).
In addition, the three branches L, C and U can be unfolded and conveniently visualised on a two-dimensional plane spanned by the coordinate $l$, the modal structure index,
and $\eta = m - n$, representing the distance in modal space between pairs of temporal wavenumbers. This unfolding process is shown in panel (b) of figure \ref{fig:Nijk_sum}, and when repeated for all modal structures leads to the distribution shown in figure \ref{fig:Energy_cuts}-(a). 
In figure \ref{fig:Energy_cuts}-(b), we report the average transfer rate $\hat{\mathsf{N}}_{lmn}$ normalised with the total average transfer rate for each structure, the quantity 
\begin{equation}\label{eq:Ti_equation}
    \hat{T}_l = \sum_{m=1}^{N/2} \sum_{n=1}^{N/2} \hat{\mathsf{N}}_{lmn}, \quad l = 1,\ldots, N/2,
\end{equation}
to illustrate more clearly the relative strength of the interactions. In figure \ref{fig:Energy_cuts}-(c), the normalised mean transfer rate for $l=1$ and 150 is reported. Interactions between triads of pairs of DFT modes are organised in agreement with the physics of scale interactions previously discussed for POD models. In absolute terms, the most relevant interactions are clearly those located near the origin of the coordinates. These correspond to low-index modes where nonlinear interactions with other low-index modes dominate, while interactions with the high-index modes, for larger $\eta$, are less important. This suggests that a sparsification approach based on pruning the interactions involving the high-index modes, i.e. the small scales, would be effective.  By contrast, for high-index modes, relevant energy interactions are organised in bands along the the axes $m$ and $n$ and involve energy exchange between low-index modes and high-energy, high-index modes. This suggests that the dynamics of the small scales is driven primarily by non-local interactions with the largest structures of the flow and not by small-scale/small-scale quadratic interactions. The slight asymmetry visible in panel (c) arises from the structure of the coefficients tensor $\mathsf{Q}_{ijk}$ and has the same physical origin as that observed in figure \ref{fig:N_ijk_POD} for the POD model.


\subsection{Sparsification of POD-based models}\label{sec:sparsification_POD}
We now apply the methodology presented in section \ref{sec:methodology} to three POD-based models resolving $90\%$, $95\%$ and $99\%$ of the kinetic energy, respectively (see Table \ref{tab:kd} for details).
Because the size of the database matrix ${\boldsymbol \Theta}(\mathsfbi{A})$ grows quadratically with the number of modes, the number of possible interactions $q$ can easily become larger than the number of available snapshots $N_t$, resulting in an underdetermined regression problem and overfitting. This is a well understood issue in data analysis and requires cross validation techniques to ensure the statistical reliability of the result \citep{stat_learning}. In this work, we employed K$-$fold cross validation, using typically $\mathrm{K}=10$. Briefly, the database is first divided into $\mathrm{K}$ folds. The model is trained using $\mathrm{K}-1$ blocks and the reconstruction error $\epsilon$ of equation (\ref{eq:epsilon}) is obtained from the fold that was left out. This procedure is iterated over all folds, obtaining the mean and the standard deviation of $\epsilon$.

\begin{figure}
  \centerline{\includegraphics[width=1\textwidth]{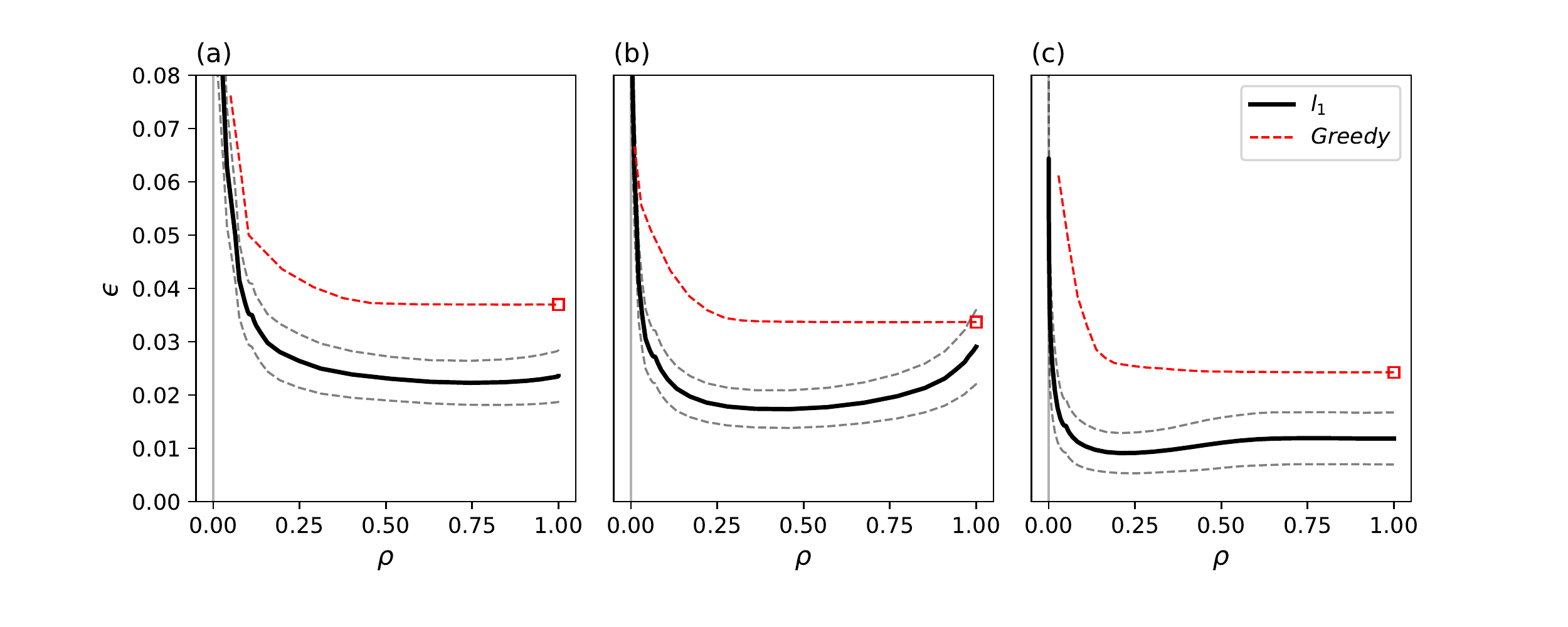}}
  \caption{$\rho-\epsilon$ curves for three POD models resolving 90\%, 95\% and 99\% of the kinetic energy in panels (a), (b) and (c), respectively. The black line represents the cross-validated error averaged over $\mathrm{K}=10$ folds. The dashed grey lines represent plus/minus one standard deviation of the cross-validated error calculated over the folds. \textcolor{black}{The red dashed line shows the reconstruction error obtained with the greedy approach}. The squares indicate the global reconstruction error of the Galerkin model obtained directly from projection. }
  \label{fig:rho_eps_POD}
\end{figure}

Figures \ref{fig:rho_eps_POD}-(a,b,c) show the sparsification curves on the $\rho-\epsilon$ plane for the three POD models considered. The mean of $\epsilon$ across the folds is displayed as a thick black line, while the grey dashed line indicates plus or minus one standard deviation. These curves have been obtained by solving problem (\ref{eq:lasso}) using strategy S1 and progressively increasing the regularisation weight. When low weights are used, dense systems with good prediction accuracy are obtained. The opposite is true for large weights, identifying points in the left part of the graph characterised by low density and poor prediction accuracy. As postulated in section \ref{sec:methodology}, the curves show a sweet spot at around $\rho \approx 0.2$, displaying a plateau for $\rho \gtrsim 0.2$, while the error $\epsilon$ grows quickly when $\rho \lesssim 0.2$. These results indicate that it is possible to prune about 80\% of the quadratic interactions in model (\ref{eq:galerkin}) without influencing the average prediction accuracy.  \textcolor{black}{The red dashed line represents the reconstruction error obtained with a naive sparsification approach. The approach consists in pruning coefficients of $\mathsf{Q}_{ijk}$ in the area denoted as $\mathrm{HH}$ in figure \ref{fig:N_ijk_POD}-(a).  The approach exploits the structure of $\mathsf{N}_{ijk}$ and is therefore referred to as ``greedy''. By varying $n \in (1, N)$, models with different sparsity are obtained, with $n=N$ corresponding to the original projection model projection (indicated as a red square in figure \ref{fig:rho_eps_POD}). The shape of the sparsification curves for the greedy approach are similar to those obtained with the $l_1$ regression. This is a direct consequence of the existence of a subset of most relevant energy interactions. However, the reconstruction error obtained from the greedy method is generally higher than that obtained by the $l_1$ regression, since the optimisation procedure involved in the $l_1$ approach modulates the strength of the remaining interactions by tuning the active quadratic coefficients, minimising the prediction error. As we show later in section \ref{sec:time_int} dedicated to analysing the model performance in time integration, this difference will have a marked effect on the long-term temporal stability of the models.}

\textcolor{black}{Regardless of the approach, the mean reconstruction error decreases as the resolved energy increases, moving from panel (a) to panel (c), as more modes participate in capturing the dynamics of velocity fluctuations.} In addition, larger models can be more effectively sparsified, as the sparsification curve drops more rapidly. This results from the non-local structure of energy interactions shown in figure \ref{fig:N_ijk_POD}. When one additional low-energy mode is included, the number of relevant interactions to be retained in the model \textcolor{black}{only grows as $\mathcal{O}(N)$ and not as $\mathcal{O}(N^2)$}, i.e. all non-local interactions with the rest of the hierarchy denoted as $\mathrm{LL}, \mathrm{LH}$ and $\mathrm{HL}$ in figure \ref{fig:N_ijk_POD}-(a). Since the total number of possible interactions grows as $\mathcal{O}(N^3)$, larger models can be more effectively sparsified. This is conceptually in agreement with the observations of \citet{taira2016network} on the sparsification properties of discrete vortex models. We also observe that the mean prediction error does not necessarily decrease monotonically when the density increases. This phenomenon is particularly visible for the model in panel (b) but all models reproduce the same behaviour. This is a symptom that the number of available snapshots (1500) is potentially not large enough for the number of coefficients ($q = N \times (N+1) / 2 + N+1 = 2926$ for the model in panel (c)) and overfitting would have occurred if no cross-validation had been performed.

\subsection{\textcolor{black}{Energy interactions identified by the regression and conservation properties}}
\begin{figure}
  \centerline{\includegraphics[width=1\textwidth]{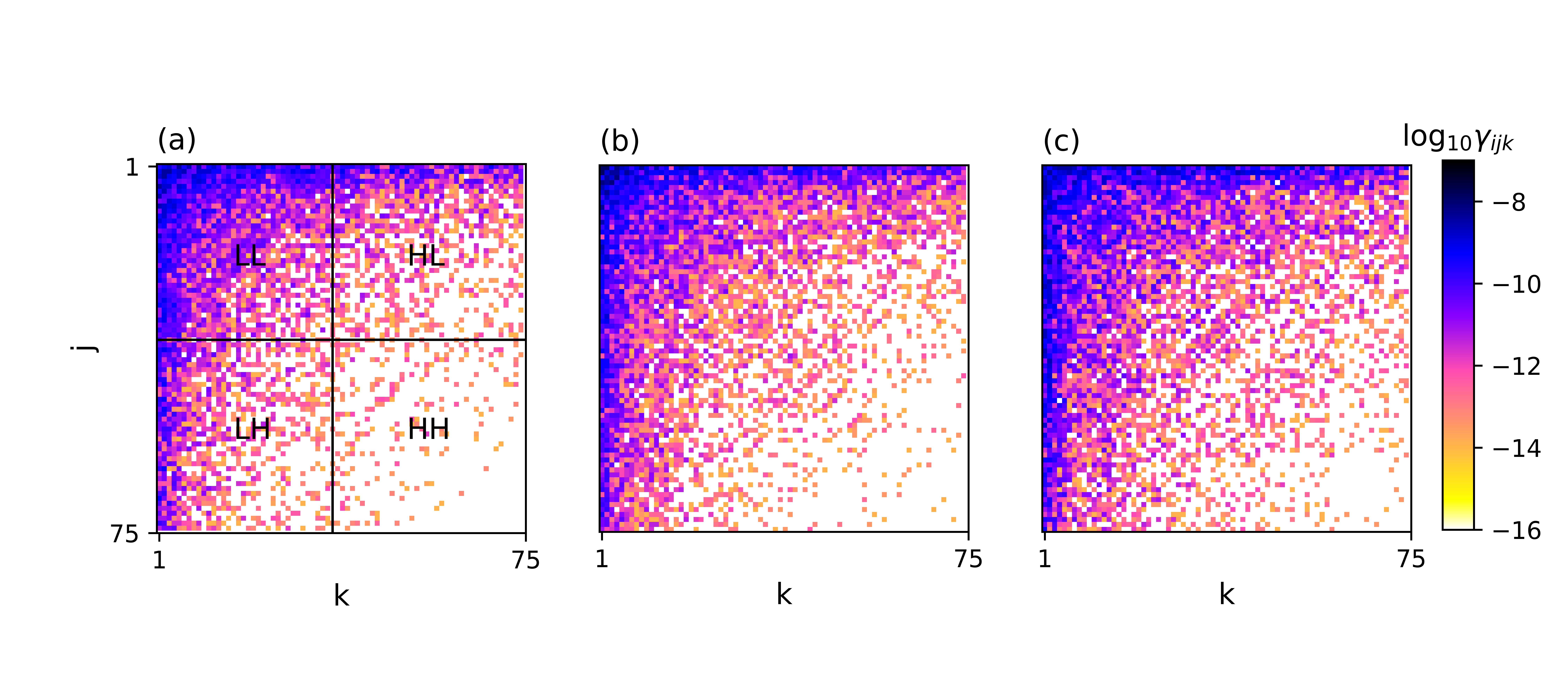}}
  \caption{Distribution of the base ten logarithm of $\gamma_{ijk}$ for $i=1,10$ and $75$, in panels (a), (b) and (c), respectively, for the POD model resolving 99\% of the fluctuation kinetic energy.}
  \label{fig:gamma_ijk_POD}
\end{figure}
To visualise the sparsity pattern identified by the regression as the regularisation weight in equation (\ref{eq:lasso}) is increased \textcolor{black}{(constant for all modes in strategy S1)}, we introduce the tensor $\boldsymbol {\gamma}$ with entries $\gamma_{ijk}$ defined as the value of the regularisation weight at which the corresponding coefficient $\mathsf{Q}_{ijk}$ is shrunk to zero by the LASSO. Figures \ref{fig:gamma_ijk_POD}-(a,b,c) show three slices of $\boldsymbol{\gamma}$ for modes $i=1, 10$ and $75$, respectively, for the largest POD model considered, \textcolor{black}{capturing $99\%$ of the total fluctuation kinetic energy}. The first interactions to disappear are the small-scale/small-scale interactions. Increasing the penalisation, interactions that are local in modal space are progressively pruned, leaving only non-local interactions involving triadic exchanges with the low-index modes for large penalisations. Interestingly, this pattern does not change qualitatively nor quantitatively as the modal index $i$ increases. In fact, a comparable number of interactions is retained across the hierarchy and the governing equations of all modes are sparsified by an equal amount. Hence, sparsification has not produced mode truncation, which would have occurred if all coefficients of some low-energy modes had been shrunk to zero by the LASSO. This behaviour can be justified by noting that the mean \textcolor{black}{square} acceleration $|| \dot{\mathsfbi{A}}_i||_2^2$ of the POD \textcolor{black}{amplitude coefficients} varies \textcolor{black}{weakly} with $i$. In fact, the sparsification pattern does not change significantly when strategy S2 is used.
\textcolor{black}{In figures \ref{fig:N_ijk_sparse}-(a,b,c) the base ten logarithm of the mean energy interaction tensor $\mathsf{N}_{ijk}^s$ computed as in \eqref{eq:non_lin_int} with the sparse coefficient tensor $\mathsfbi{Q}^s$ is shown for the same three modal indices as in figure \ref{fig:gamma_ijk_POD}. The data refer to sparse models with $\rho=0.3$, located nearby the sweet spot of the curves in figure \ref{fig:rho_eps_POD}-(c).  
\begin{figure}
  \centerline{\includegraphics[width=1\textwidth]{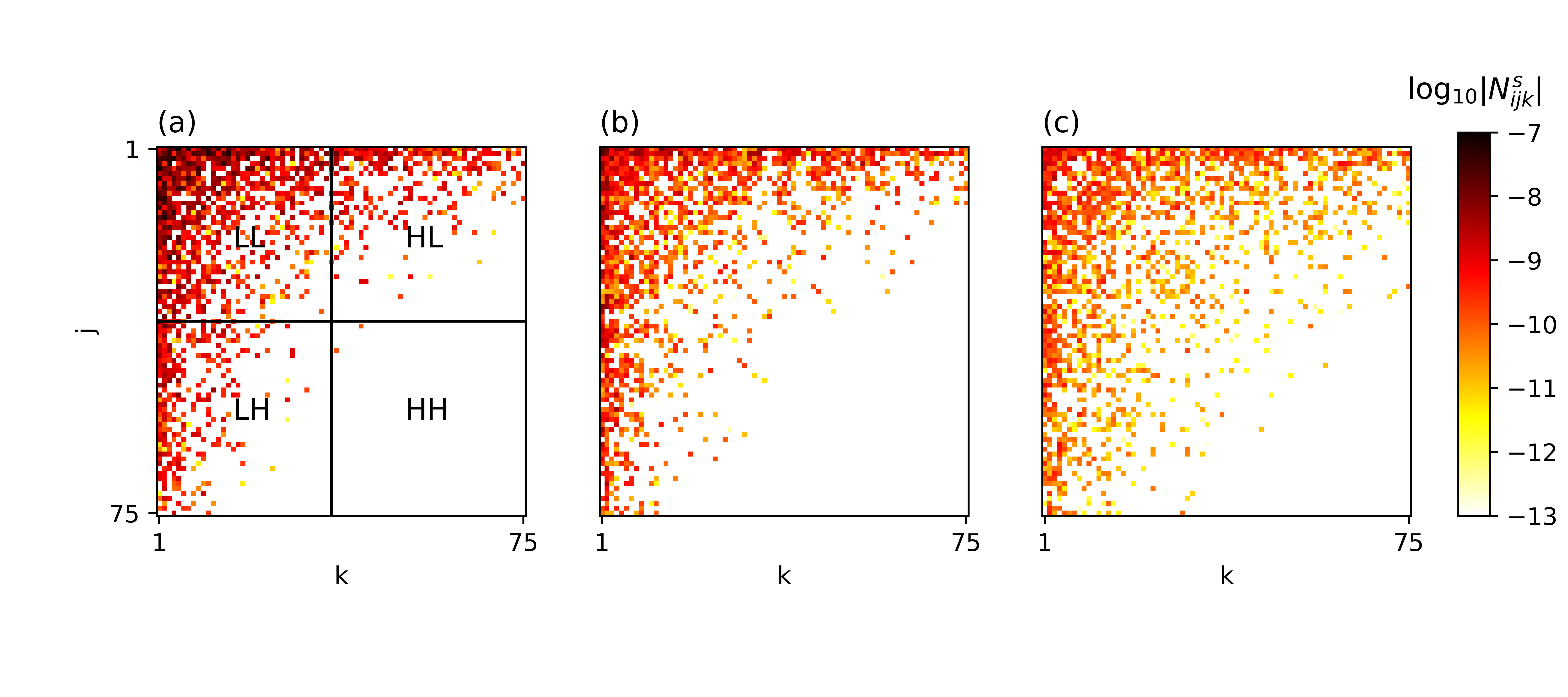}}
  \caption{Base ten logarithm of the sparsified interaction tensor $\mathsf{N}^s_{ijk}$ for $i=1,10$ and $75$ in panels (a), (b) and (c), respectively, for a POD model model resolving $99\%$ of the fluctuation kinetic energy and  $\rho = 0.3$.}
  \label{fig:N_ijk_sparse}
\end{figure}
It can be observed that the sparsified model has a pattern of interactions resembling that of the dense model in figure \ref{fig:N_ijk_POD}. However, weak interactions and the associated flow physics have been pruned.  It is also clear that the asymmetry of the interaction pattern observed in figure \ref{fig:N_ijk_POD} and the physical mechanism that originates it are invisible to the regression and the interaction pattern in figure \ref{fig:N_ijk_sparse} is now symmetric with respect to a swap of the indices $j, k$.} 

\begin{figure}
  \centerline{\includegraphics[width=0.9\textwidth]{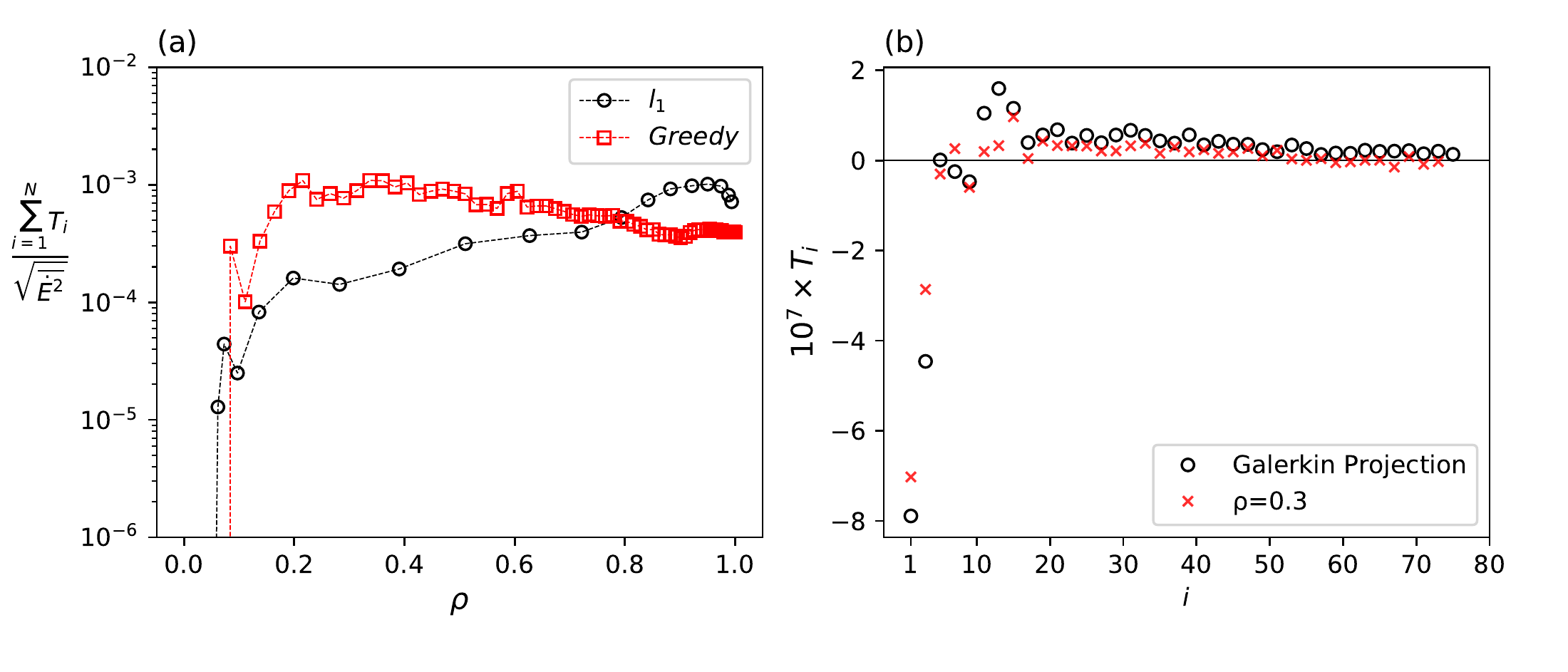}}
  \caption{
 \textcolor{black}{Panel (a): the normalised energy conservation error as a function of the density $\rho$, for the greedy sparsification approach (empty squares) and the $l_1$ based approach (empty circles).}  Panel (b): the total net energy transfer rate $T_i$ as a function of the modal index $i$ for two POD models resolving $99\%$ of the kinetic energy, with coefficients identified from projection and for a $l_1$ sparse model. One every two data points is reported. }
  \label{fig:Ti_pod}
\end{figure}

\textcolor{black}{Despite the aggressive pruning, the sparse models reproduce fairly accurately the overall structure of the intermodal energy budgets. In the present flow configuration, the convective nonlinearity is energy conserving and Galerkin models should obey the relation  $\sum_{i=1}^N T_i = 0$ as $N\rightarrow\infty$, with \mbox{$T_i = \sum_{j=1}^{N}\sum_{k=1}^{N} \mathsf{N}_{ijk}$} the time averaged energy transfer rate to/from mode $i$. For finite-dimensional approximations, this property is not satisfied exactly and the residual of the summation can be taken as a measure of the overall energy balance. Figure \ref{fig:Ti_pod}-(a) shows such residual for the $l_1$ sparsified models (empty circles) and for the models obtained with the greedy approach (empty squares), as a function of the density. The residual is normalised by the root mean square value of the rate of change of the integral fluctuation energy. Note that the greedy model at $\rho=1$ is the model obtained directly from projection. It can be observed that the energy conservation error is relatively small, in the order of $10^{-3} \div 10^{-4}$. For large densities, it is larger than that of the projection model, because the regression tunes model coefficients to minimise the mean square error on the modal accelerations and does not enforce this physical constraint directly. The energy conservation residual decreases for sparser models and is ten times smaller than the projection model, owing to the lower number of active coefficients that participate in the regression. Figure \ref{fig:Ti_pod}-(b) shows the distribution of the time averaged energy transfer rate associated to mode $i$ for the $l_1$ sparsified model at $\rho=0.3$ (red crosses) and the dense model obtained from projection (empty circles). Data is reported every second mode. For the projection model, the net energy transfer is negative for the first few modes and changes sign at $i \sim 10$. Physically, this trend suggests that the first few modes extract energy from the mean flow and feed the dissipative high-index modes via triadic interactions. The $l_1$ sparsified model correctly reproduces this global trend,}
\textcolor{black}{even though no constraints have been introduced \citep{loiseau2018constrained}.} As discussed in section \ref{sec:methodology},  it is argued that this is a general properties of data-driven techniques relying on optimisation ideas, such as the LASSO, which naturally reproduce invariants and conservation properties embedded in the data to a level defined by noise levels. For instance, \citet{taira2016network} used network-theoretic ideas to sparsify connections in a discrete vortex model and observed that sparsification conserves the invariants of discrete vortex dynamics.



\subsection{\textcolor{black}{Long-term temporal behaviour of the $l_1$ sparse systems}}\label{sec:time_int}
\begin{figure}
  \centerline{\includegraphics[width=1.05\textwidth]{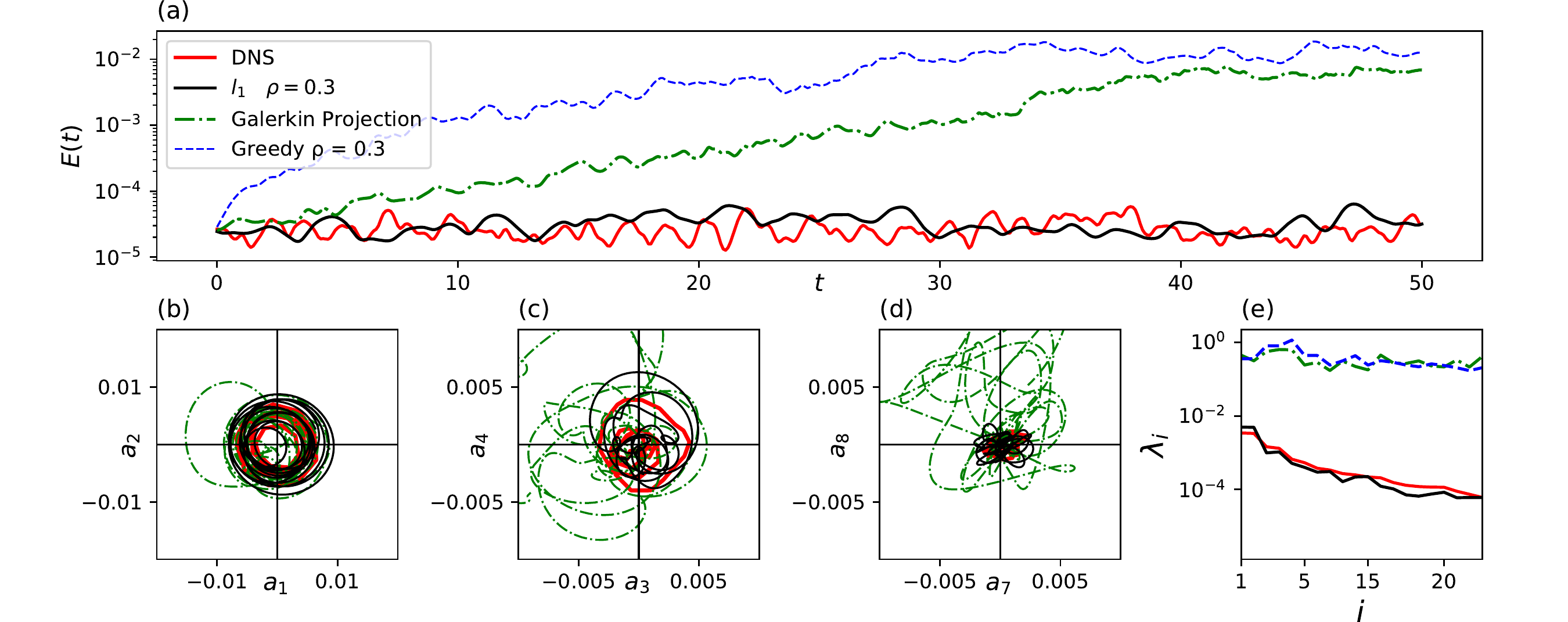}}
  \caption{\textcolor{black}{Panel (a): temporal evolution of the turbulent kinetic energy $E(t)$ from DNS compared to that obtained from temporal integration of the $l_1$, greedy and Galerkin projection models. Sparse models have $\rho = 0.3$. Panels (b) to (d): state space projections onto three different mode pairs. Data for the greedy model is omitted as the trajectory quickly leaves the visible range.  Panel (e): average modal energy ($\lambda_i  = \overline{a_ia_i}$) from DNS and long-time integration of the models considered in panel (a).}}
  \label{fig:POD_sparse_timeint}
\end{figure}

\textcolor{black}{We now turn our attention to the long-time behaviour of the sparsified models under temporal integration. We consider results for models resolving $95\%$ of the turbulent kinetic energy as an illustrative example, and use the projections of the POD modes onto one of the DNS snapshots to obtain initial conditions. Results are shown in figure \ref{fig:POD_sparse_timeint}}.
\textcolor{black}{Panel-(a) shows the temporal evolution of the integral fluctuation kinetic energy $E(t)$, as defined in \eqref{eq:energy_POD}, for the dense projection model and the sparse models obtained with the $l_1$ and greedy approach, with same density  $\rho = 0.3$. The temporal evolution obtained with these models is compared against the fluctuation kinetic energy from DNS. For the projection model, the integral kinetic energy grows substantially in the first 40 time units and then saturates on a value that is about two orders of magnitude larger than what observed in DNS. The over-prediction occurs because the truncation of the small scales in the ansatz (\ref{eq:modal_expansion}) leads to a significant imbalance of the production--dissipation budget within the model \citep{balajewicz2013low,noack2005need,noack2008finite,noack2011reduced}. A qualitatively similar behaviour, if not worse, is then necessarily observed for the model sparsified with the greedy approach, since neglecting weak interactions alone does not cure the original dissipation problems. Conversely, the $l_1$ model is able to predict the correct average fluctuation kinetic energy and has excellent long-term stability properties, despite this being not enforced in the regression procedure (see \citet{fick2017reduced}). We argue that this is due to the fact that the $l_1$ procedure performs a ``prune-then-calibrate'' approach, where weak interactions are first pruned and the remaining active coefficients are then tuned in the optimisation involved in (\ref{eq:lasso}) to match the reference dynamics. It is evident from these results that this second step is key to obtain accurate long-term behaviour. Panels (b) to (d) of figure \ref{fig:POD_sparse_timeint} show a shorter segment of state space trajectory from these models projected onto three different pairs of modes. We omit the data for greedy approach since orbits quickly drift out of the visible range and is thus qualitatively similar to that from the dense model. It is clear that the trajectory of the $l_1$ model remains in the same volume of state space occupied by the DNS projections, while the projection model drifts away to a different region of state-space, over-predicting the integral kinetic energy. This is more effectively visualised in panel (e), displaying the average modal energy $\lambda_i  = \overline{a_ia_i}$ as a function of the modal index. Data is obtained from averaging long trajectories after the initial transient in panel (a) has completed. The projection and greedy models predict much larger energies across the entire spectrum, while the $l_1$ correctly predicts the correct decay of modal energies. It is, of course, not possible to guarantee that $l_1$ sparsified models of generic turbulent flows will have good long-term stability \citep{schlegel_noack_2015}, but the present results constitute evidence that this is realistically possible on a non-trivial problem. Finally, we have observed in animations of the reconstructed flow fields using the spatial modes and the temporal coefficients from numerical integration (see supplementary material) that characteristic flow features, such as the erratic burst of corner vortices and the evolution of coherent structures in the shear layer bounding the main vortex, are also well reproduced by the $l_1$ model, providing a realistic flow reconstruction over long time horizons.}

\subsection{Sparsification set up for DFT-based models}
\label{sec:sparsification_DFT}
Before moving to the sparsification of DFT-based \textcolor{black}{models}, we briefly discuss three technicalities arising from the oscillatory nature of the temporal coefficients. 
\begin{figure}
  \centerline{\includegraphics[width=0.9\textwidth]{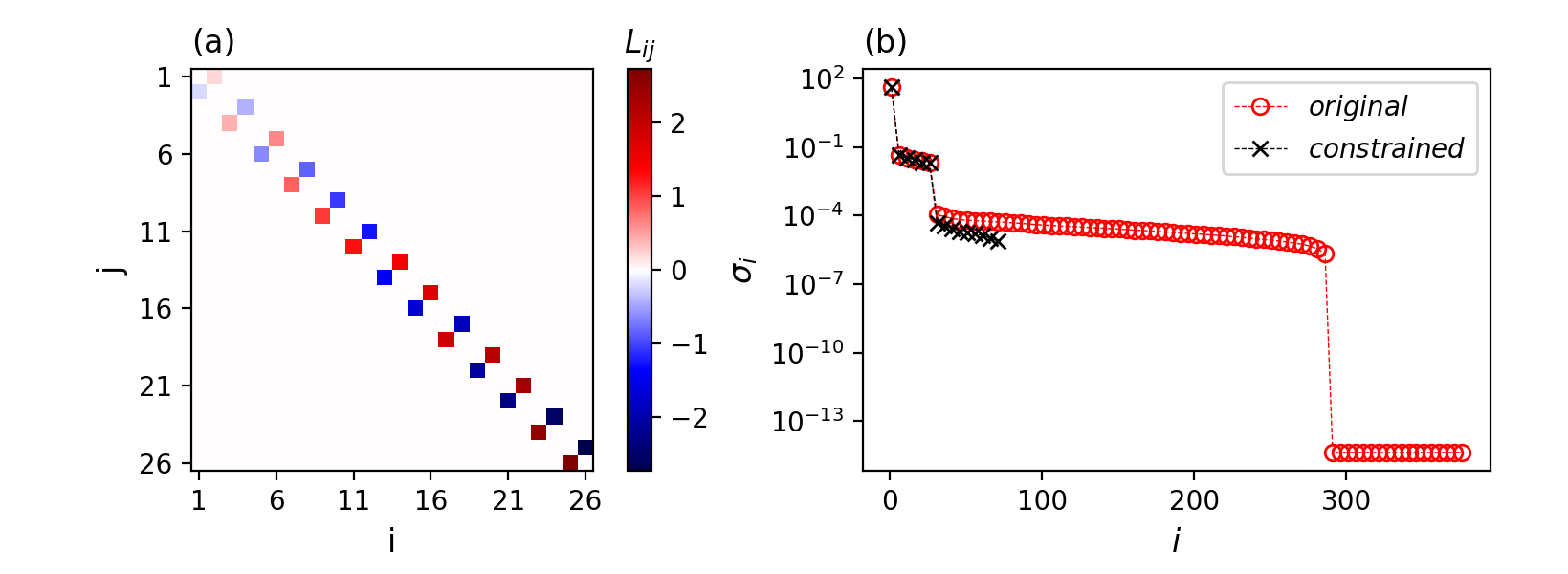}}
  \caption{\textcolor{black}{Panel (a): entries of the linear tensor $\mathsfbi{L}$ identified by the unconstrained regression. Panel (b): singular values $\sigma_i$ of the full database matrix $\boldsymbol{\Theta}(\mathbf{A})$ of equation \eqref{eq:database} (red circles), and of the reduced matrix (black crosses) obtained by keeping only the subset of columns corresponding to active interactions on the three branches of figure \ref{fig:Nijk_sum}. One every five singular values is shown for clarity.}}
   \label{fig:Lin_part}
\end{figure}
As an illustrative example, we consider a small-sized model constructed with \mbox{$N=26$} modes and perform sparsification as discussed in section \ref{sec:methodology}, with a relatively small regularisation weight ($\gamma = 10^{-14}$, \textcolor{black}{strategy S1}).
The first key result is that all the entries of the constant and quadratic coefficient tensors $\mathsfbi{C}$ and $\mathsfbi{Q}$ are set to zero, while the linear tensor $\mathsfbi{L}$ has a characteristic bidiagonal structure, shown in figure \ref{fig:Lin_part}-(a). The system identified by the regression is equivalent to a set of $N/2$ decoupled linear oscillators in the form 
\begin{equation}
\left[ \begin{array}{c} \dot{a}_{2l-1} \\ \dot{a}_{2l} \end{array} \right] = \omega_l \begin{bmatrix} 0 & 1 \\ -1& 0 \end{bmatrix}  \left[ \begin{array}{c} a_{2l-1} \\ a_{2l} \end{array} \right] \quad \quad l = 1,..., N/2,
\end{equation}
coupling pairs of temporal coefficient oscillating at the same angular frequency \mbox{$\omega_l = 2\pi l / T$}, with $T$ being the observation time. The eigendecomposition of the tensor $\mathsfbi{L}$ is trivial. Eigenvalues are all imaginary and come in pairs that are integer multiples of the fundamental frequency $\omega_1 = 2 \pi/T$. 
While this result is consistent with recent ideas on Koopman operator theory \citep{mezic2013analysis}, where nonlinear dynamics are modelled with a linear system of larger dimension, all information on nonlinear energetic interactions has been lost in the process since the nonlinear part of the system has been completely eliminated by the regression. This result is due to the fact that, when temporal coefficients are sine/cosine pairs, there is a column of ${\boldsymbol \Theta}(\mathsfbi{A})$ that is exactly parallel to the target $\dot{\mathsfbi{A}}_i$, since time differentiation is equivalent to a permutation of sine/cosine pairs. As pointed out in \citet{brunton2019machine} incorporating and enforcing known flow physics is a challenge and opportunity for machine learning algorithms. In order to address this first aspect, we introduce a physically-motivated approach based on considerations of the time averaged energy budget of system (\ref{eq:energy_POD}). Since the temporal coefficients have zero mean and are uncorrelated in time, we should obtain
\begin{equation}\label{eq:energy_avg}
    \mathsf{L}_{ii}\overline{a_i a_i} + \sum_{j=1}^N \sum_{k=1}^N \mathsf{Q}_{ijk}\overline{a_ia_ja_k} = 0, \quad i = 1, \ldots, N
\end{equation}
i.e. only the diagonal element of the linear term participates in the mean power budget. Hence, for the sparsification of DFT-based models we use a modified database matrix that only contains the column associated to the diagonal part of the linear term. 


The second aspect is that for DFT models the database matrix $\boldsymbol{\Theta}(\mathsfbi{A})$ is not full rank and some of the columns of this matrix are linearly dependent. In this case, the LASSO is known to select one column at random (according to the particular ordering of the columns) and sets to zero regression coefficients of the other linearly dependent columns \citep{tibshirani2013lasso,hastie2015statistical}. Machine learning techniques often come without guarantees for robustness \citep{brunton2019machine}, implying that physical insight obtained with these tools might be questionable. \textcolor{black}{To avoid this problem, we constructed a reduced database matrix ${\boldsymbol \Theta}(\mathsfbi{A})$ containing only columns corresponding to the interactions on the three branches of figure \ref{fig:Nijk_sum}.}
The reduced database matrix is full rank, as can be seen in panel (b) of figure \ref{fig:Lin_part}, showing the singular values of the full database matrix defined by equation (\ref{eq:database}) and of the reduced matrix. The important consequence is that the solution of the LASSO problem (\ref{eq:lasso}) is unique \citep{tibshirani1996regression}, and can be thus compared with the available physical knowledge of scale interactions in turbulent flows. \textcolor{black}{This reduction is not strictly necessary, since the $l_1$ regression identifies this pattern anyway for fairly small regularisation weights. However, this has the advantage that} the computational complexity of sparsifying the entire Galerkin model only grows as $\mathcal{O}(N^2)$ instead of $\mathcal{O}(N^3)$, as for POD models, because the reduced database matrix contains a number of interactions equal to $q = 2(N+1)$ at most. More importantly, \textcolor{black}{because of the greatly  reduced number of free coefficients} cross-validation techniques to avoid over-fitting become unnecessary.

The third aspect of DFT-based models is that, as anticipated, the number of modes is not uniquely defined by the energy resolution but depends on the overall observation time. Long observation times would be beneficial to reach statistical significance but would result in low-energy/low-frequency modes that do not contribute significantly to the overall dynamics. In practice, we have divided the original dataset into $M$ partitions and performed DFT for each of them separately. Then, we stacked vertically the modal acceleration matrices and the reduced database matrices from the partitions and solved \eqref{eq:lasso} for a common coefficients vector.



\subsection{Sparsification of DFT-based models}
\textcolor{black}{We now move to the sparsification of DFT-based models. We focus primarily on the structure of energy interactions identified by the regression and leave long-term temporal stability considerations aside. In fact, the DFT produces modal structures by assuming \textit{a priori} their temporal behaviour, i.e. harmonic motion, and the meaning of a time-domain analysis is thus conceptually unclear.}

\begin{figure}
  \centerline{\includegraphics[width=0.95\textwidth]{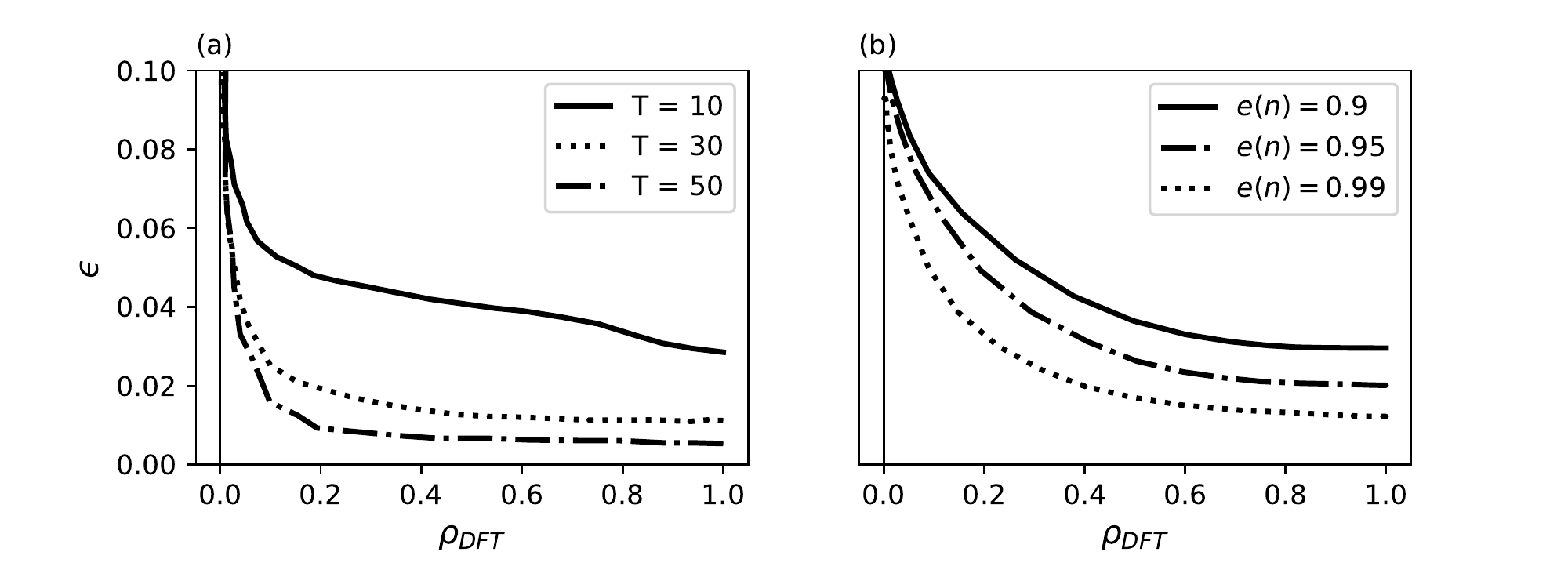}}
  \caption{Panel (a): sparsification curves for models obtained by three different observation times and resolving $100\%$ of the kinetic energy. Panel (b): sparsification performed with $T=30$ with three different energy resolutions $e(n)$. \label{fig:rho_epsilon_DFT}}
\end{figure}

\textcolor{black}{We} introduce the modified density $\rho_{\mathrm{DFT}}$ spanning the range $[0, 1]$ and representing the number of active coefficients with respect to the total number of active interactions on the three branches of figure \ref{fig:Nijk_sum}. For large models, the approximation $\rho_{\mathrm{DFT}} \approx 2/3\rho N$ can be used. In figure \ref{fig:rho_epsilon_DFT}-(a), sparsification curves for three models obtained with observation times $T=10$, $30$ and $50$ (with $M=15$, $5$ and $3$ partitions of the full dataset, respectively), at full energy resolution, are reported. Strategy S1, where the regularisation weight is maintained constant for all modes is used. We observe that the error decreases monotonically with the observation time.
This is a consequence of the larger number of frequencies that interact quadratically to reconstruct the original DNS acceleration data. For the larger model obtained at $T=50$, $70\%$ of the triadic interactions can be pruned with no major effects on the overall prediction error. If the full coefficient tensor $\mathsfbi{Q}$ is considered, this correspond to a remarkably low density of 0.0015.
Figure \ref{fig:rho_epsilon_DFT}-(b) shows the sparsification curves for models obtained with observation time $T = 30$, for three different energy resolutions, $e(n) = 0.9$, $0.95$ and $0.99$. Interestingly, we notice that the curves do not present a plateau for high densities as opposed to the full resolution mode shown in panel (a) and the POD sparsification curves of figure \ref{fig:rho_eps_POD}. This is the combined effect of the dramatic decrease in the number of modes at lower energy resolutions (see table \ref{tab:kd}) and the inherent efficient description of energy interactions in DFT-based models compared to POD. 

\begin{figure}
   \centerline{\includegraphics[width=1\textwidth]{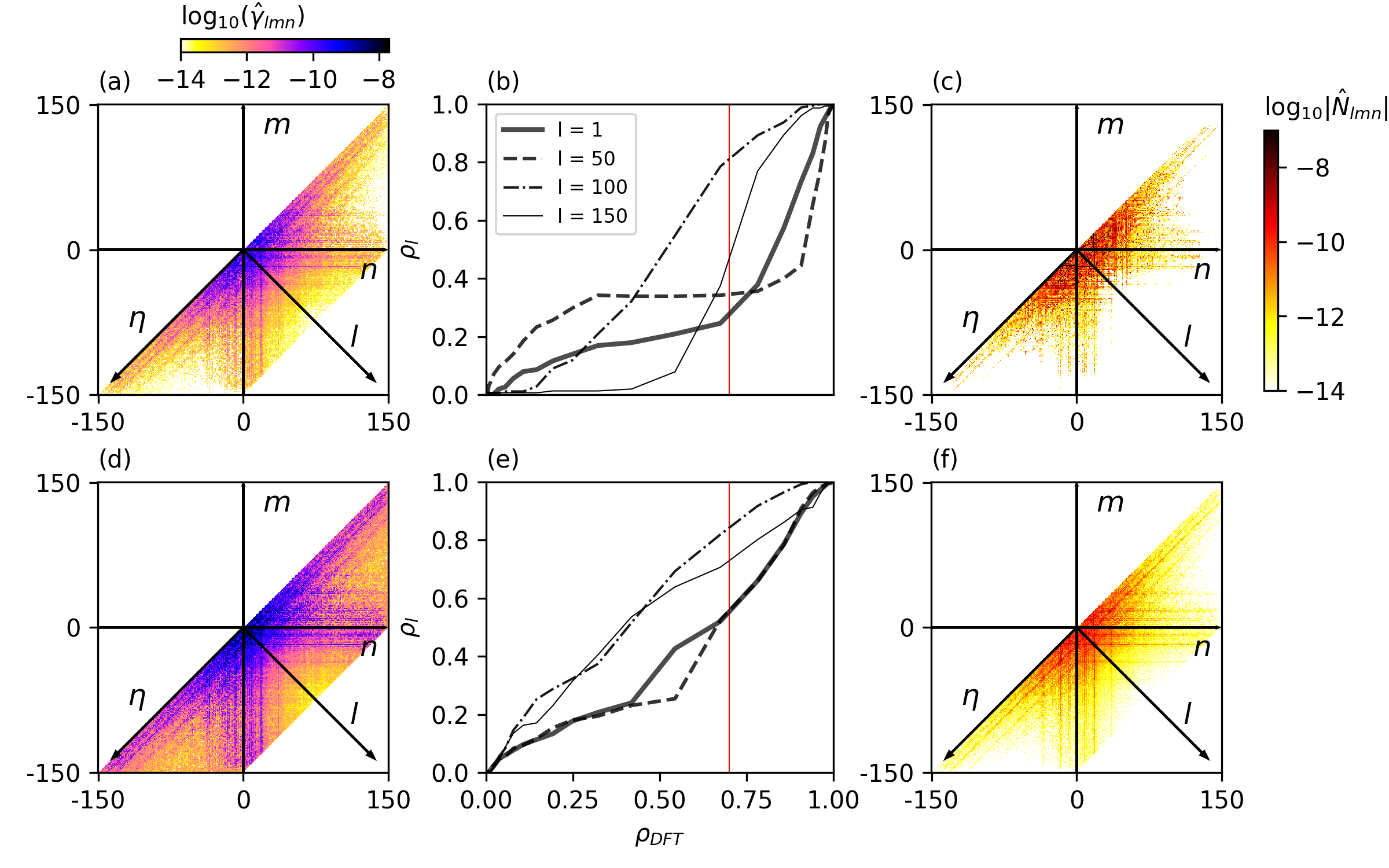}}
 \caption{Top panels: strategy S1; bottom panels: strategy S2. Panels (a) and (d) show the distribution of $\hat{\boldsymbol{\gamma}}$. Panels (b) and (e) show the trend of the modal density $\rho_l$ against the global density $\rho_{\mathrm{DFT}}$ for four different modes in different parts of the spectrum. Panels (c) and (f) show the energy interaction tensor $\mathsfbi{N}^s$ of the sparsified system. \label{fig:gamma_ijk_DFT}}
\end{figure}

We now compare strategies S1 and S2 on the full resolution model obtained with observation time $T=30$.
Results of this analysis are reported in figure \ref{fig:gamma_ijk_DFT}. The top/bottom panels are obtained with the strategy S1/S2. Panel (a) shows the tensor $\hat{\boldsymbol{\gamma}}$, obtained by processing and visualising the full tensor $\boldsymbol{\gamma}$ using the same technique utilised for the interaction tensor $\hat{\mathsfbi{N}}$ in figure \ref{fig:Nijk_sum}. Panel (b) shows the density of individual ordinary differential equations for a selected number of modal structures as a function of the overall model density $\rho_{\mathrm{DFT}}$, while the sparsified interaction tensor $\hat{\mathsfbi{N}}^s$ for $\rho_{\mathrm{DFT}} = 0.7$ is shown in panel (c).
When the regularisation weight is maintained constant, the sparsification pattern emerging from the tensor $\hat{\boldsymbol{\gamma}}$ follows the distribution of the mean energy transfer rate of figure \ref{fig:Energy_cuts}-(a). In particular, despite the signature of non-locality is still visible in the pattern, the sparsification is highly skewed across the spectrum because the equations for high-frequency modes are excessively sparsified for moderate penalisations as opposed to those of low-frequency, high-energy modes. 

This behaviour is better seen in the individual density curves in panel (b). Specifically, the density $\rho_l$ of the last mode pair ($l=150$) drops quite pronouncedly to much lower density than average at $\rho_{\mathrm{DFT}} \approx 0.5$.  Panel (d) shows the sparsification pattern obtained with the second strategy. We observe that, in this case, the interactions are retained according to their relative strength producing a sparsification pattern that follows the relative energy transfer rate reported in figure \ref{fig:Energy_cuts}-(b). This results in a more balanced sparsification across the spectrum, \textcolor{black}{where} the modal density $\rho_l$ decreases more uniformly for all modes as the global density is decreased, as shown in panel (e). The mean energy transfer rate of the models sparsified using the two strategies, with $\rho_{\mathrm{DFT}} = 0.7$, is reported in panels (c) and (f).  Globally, the structure and intensity of energy interactions is preserved by the LASSO, although strategy S1 has more aggressively sparsified the high-index modes and truncated the equations of the last five pairs of modes. \textcolor{black}{Although not shown here, all DFT models, regardless the strategy, have similar energy conservation properties as for POD modes, as illustrated in figure \ref{fig:Ti_pod}}.

As a final remark, we have observed in sparsification of larger DFT models that, although the LASSO is able to successfully identify the dominant subset of energy interactions, the complexity of the optimisation problem makes an accurate reconstruction of the numerical values of the system coefficients challenging. This is due to the spectral properties of the database matrix $\boldsymbol{\Theta}$ which deteriorate as the number of modes considered grows \citep{cordier2010calibration}. A potential solution to this issue would be to use elastic-net regression \citep{stat_learning} which combines an $l_1$ term with an $l_2$ (Tikhonov) penalisation. This would provide a better trade-off between sparsification and stability of the reconstructed coefficients.

\section{Concluding remarks}\label{sec:conclusions}

In this paper, we applied recent \textcolor{black}{data-driven} methods for the identification of sparse dynamical systems to sparsify nonlinear triadic interactions in projection-based reduced order models of \textcolor{black}{two-dimensional unsteady} flows. Our work is motivated by established knowledge of scale interactions in turbulence, whereby dynamics at a certain length scale depend most prominently on a subset of other length scales. Computationally, our methodology is based on $l_1$-based regression methods and is scalable to large models defined by hundreds of modal structures. These methods are used to recast the problem of identifying relevant triadic interactions into a convex optimisation problem for which scalable, efficient solvers can be used. 
The overarching aim is to develop large reduced order models covering a wide range of length scales, but where computational efficiency and physical interpretability have been preserved by pruning weak triadic interactions. 

In this analysis we considered two-dimensional lid-driven cavity flow at Reynolds number $Re = 2 \times 10^4$. We generated two families of reduced order models by Galerkin projection of the Navier-Stokes equations onto the subspace spanned by Proper Orthogonal Decomposition and Discrete Fourier Transform modes. The goal was to understand the role of the subspace utilised for projection on the structure and sparsity of energy interactions between modes. As discussed in \cite{brunton2019machine} an open problem in applying machine learning algorithms to fluids problems is to successfully incorporate known flow physics. In our case, we have observed that for DFT-based models, it has become necessary to manually modify the database matrix in order to ensure the uniqueness of the solution and preserve the full nonlinear character of \textcolor{black}{modal} interactions. The analysis of the average energy transfer rates between modal structures has shown \textcolor{black}{that, for both the POD- and DFT-based models, a subset} of most relevant interactions exists, in agreement with the established picture of scale interactions in two-dimensional flows. \textcolor{black}{This is an \textit{a-posteriori} feature of solutions of the equations and not an \textit{a-priori} property of the evolution equations. In fact, the model coefficients identified by the Galerkin projection do not have a particular structure and are typically different from zero}. Our results show that, in both cases, there exists a sweet-spot on the $\rho-\epsilon$ curve where the sparsification approach recovers 
correctly this subset\textcolor{black}{, with little effect on the prediction accuracy.} 
\textcolor{black}{The models also preserve to a good degree of accuracy the non-local nature of triadic interactions and the conservation properties of the convective term of the Navier-Stokes equations. In principle, energy conservation could be enforced exactly \citep{loiseau2018constrained}, although we have not found this to be necessary to obtain satisfactory temporal stability characteristics. In fact, unlike dense models obtained directly from the projection, the $l_1$ sparsified POD-based models have excellent long-term stability properties. Numerical integration shows that trajectories generated by these models remain in the same area of state space occupied by the DNS projections. The average integral fluctuation kinetic energy and the distribution of energy across modes are also reproduced fairly well. Sparse models constructed by a naive procedure where weak interactions are pruned, referred to as greedy approach in the paper, do not enjoy the same robustness and have worse performance than the dense models. This indicates that, once coefficients corresponding to weak interactions have been shrunk to zero by the $l_1$ penalisation, re-balancing the remaining coefficients with the least-squares term in the LASSO problem (\ref{eq:lasso}) is key to preserve accuracy.}

We have also observed that the effectiveness of the sparsification grows with the number of modes \textcolor{black}{utilised in the projection} (energy resolution). This is a result of the non-local nature of \textcolor{black}{scale interactions in two-dimensional flows}, where the dynamics of small-scale features is dominated by the advection of the large modes, rather than by \textcolor{black}{the} small-scale/small-scale nonlinearity. \textcolor{black}{The interesting consequence is that}, while the total number of \textcolor{black}{quadratic interactions grows cubically} with the number of modes, the number of relevant interactions \textcolor{black}{does not grow as quickly. Hence, our} expectation is that sparsification \textcolor{black}{becomes} more effective as the Reynolds number increases, as a result of the increased \textcolor{black}{scale separation. This would also reduce, in relative terms, the computational costs required for propagating the model forward in time. In this paper, we have not, admittedly, explored in enough detail the role of sparsification on the reduction of computational costs, as these are highly influenced by implementation details and code optimisations. For instance, the sparsified quadratic tensor $\mathsfbi{Q}^s$ could be efficiently stored and evaluated using sparse matrix techniques. Characterising physical and computational properties of $l_1$ models as a function of the Reynolds number in two- and three-dimensional turbulent flows is therefore an interesting avenue for future work.} 

A major difference between the two decompositions \textcolor{black}{considered in this work} is that energy interactions between triads of DFT modes are highly localised in modal space, \textcolor{black}{as a result of the oscillatory nature of the temporal coefficients}. On the other hand, for \textcolor{black}{energy-optimal} POD modes, temporal coefficients contain a wider range of frequencies and energy transfers are \textcolor{black}{thus} inevitably \textcolor{black}{scattered} in modal space. \textcolor{black}{Consequently}, the number of active interactions \textcolor{black}{only grows} as $\mathcal{O}(N^2)$ for \textcolor{black}{DFT models, rather than} as $\mathcal{O}(N^3)$ for \textcolor{black}{POD models. Therefore, we conclude that the sparsity of energy interactions is not necessarily invariant when analysed on different subspaces. In practice, the favourable $\mathcal{O}(N^2)$ scaling observed for DFT modes is appealing for the construction of large yet interpretable models, covering a wide range of spatio-temporal scales. However, the meaning and practical utility of temporal integration of DFT-based models (but also models constructed from other decompositions producing purely oscillatory modes), where the solution structure has been assumed a-priori, remains conceptually unclear. In summary, the stark difference between DFT- and POD-based models suggests that it might be possible to develop a modal decomposition that identifies a set of maximally independent structures, e.g. where the resulting quadratic coefficient tensor $\mathsfbi{Q}$ or the average interaction tensor $\mathsfbi{N}$ are maximally sparse. However, this appears to be a nonlinear optimisation problem, with the associated convergence and uniqueness issues. Work in this direction has been recently initiated by \citet{schmidt2020bispectral}.}

\vspace{0.5cm}

The authors gratefully acknowledge support for this work from the Air Force Office of Scientific Research (Grant No. FA9550-17-1-0324, Program Manager Dr D. Smith).






\bibliographystyle{jfm}
\bibliography{jfm-instructions}

\end{document}